\documentclass[aps,prx,twocolumn,
               superscriptaddress,
               nofootinbib,longbibliography,
               10pt,showkeys,floatfix]{revtex4-2}
\usepackage{amsmath, amssymb}
\usepackage{graphicx, nicefrac, textcomp}
\usepackage{mwe}
\usepackage{braket}
\usepackage{color}
\usepackage{dcolumn}
\usepackage[utf8]{inputenc}
\usepackage[T1]{fontenc}
\usepackage{dsfont}
\usepackage{svg}
\usepackage[version=4]{mhchem}
\usepackage[toc,page]{appendix}
\usepackage{float}
\usepackage{silence}
\newcommand{\mr}{\mathrm}

\newcommand{\mi}{\mu^i}
\newcommand{\qi}{\theta^i}
\newcommand{\oi}{\Omega^i}
\newcommand{\hi}{\Phi^i}
\newcommand{\dmi}{\Delta\mu^i}
\newcommand{\dqi}{\Delta\theta^i}

\newcommand{\al}{\alpha}
\newcommand{\ab}{{\alpha\beta}}
\newcommand{\aby}{{\alpha\beta\gamma}}
\newcommand{\abyd}{{\alpha\beta\gamma\delta}}

\newcommand{\be}{\beta}
\newcommand{\by}{{\beta\gamma}}

\newcommand{\br}{{\bf r}}

\newcommand{\bfr}{\mathbf{r}}

\newcommand{\Ga}{\alpha}
\newcommand{\Gab}{\alpha\beta}

\newcommand{\Gb}{\beta}

\newcommand{\rMM}{\mathrm{MM}}
\newcommand{\rQM}{\mathrm{QM}}
\newcommand{\rQMMM}{\mathrm{QM/MM}}

\newcommand{\rself}{\mathrm{self}}

\newcommand{\Dmu}{{\Delta\mu}}

\newcommand{\ecom}{COM}

\usepackage[colorlinks=true,citecolor=blue,linkcolor=blue,urlcolor=blue]{hyperref}

\begin{document}
\title{Polarizable Embedding QM/MM for Periodic Systems}

\author{Julian Beßner}
\affiliation{Institute of Electrochemistry, University of Ulm, 89081 Ulm, Germany}

\author{Anoop A. K. Nair}
\affiliation{Science Institute and Faculty of Physical Sciences, University of Iceland, 107 Reykjav\'ik, Iceland}

\author{Magnus A. H. Christiansen}
\affiliation{Science Institute and Faculty of Physical Sciences, University of Iceland, 107 Reykjav\'ik, Iceland}

\author{Timo Jacob}
\affiliation{Institute of Electrochemistry, University of Ulm, 89081 Ulm, Germany}
\affiliation{Helmholtz-Institute Ulm for Electrochemical Energy Storage, University of Ulm, 89081 Ulm, Germany}
\affiliation{Karlsruhe Institute of Technology, 76021 Karlsruhe, Germany}

\author{Hannes J\'onsson}
\email{hj@hi.is}
\affiliation{Science Institute and Faculty of Physical Sciences, University of Iceland, 107 Reykjav\'ik, Iceland}

\author{Elvar Örn J\'onsson}
\email{elvarorn@hi.is}
\affiliation{Science Institute and Faculty of Physical Sciences, University of Iceland, 107 Reykjav\'ik, Iceland}


\date{\today}

\begin{abstract}
 A general polarizable embedded (PE) quantum 
 mechanics/molecular mechanics scheme 
 for periodic systems is presented, 
 describing mutual polarization of the 
 two subsystems. The QM system, described 
 with density functional theory (DFT), is 
 coupled to a single center multipole 
 expansion (SCME) model, characterizing 
 \ce{H2O} molecules in the MM region. In 
 SCME the 
 \ce{H2O} molecules are ascribed
 anisotropic dipole and quadrupole 
 polarizabilities and permanent 
 multipoles up 
 to and including the hexadecapole. 
 Our embedding scheme illustrates
 a smooth and efficient convergence pattern
 of the periodic interaction potential
 by introducing a single and clustered 
 multipole expansion points in the 
 far-field.
 By choosing the near- and 
 far-field expansion of the potential 
 carefully the PE-QM/MM calculation 
 matches the level of accuracy of a 
 the QM calculation. 
 In the short range, 
 the electrostatic interaction between 
 the QM and MM subsystems is damped with a  
 real-space and pair-wise
 isotropic damping functions 
 -- resulting in a 
 screened interaction 
 and
 preventing over-polarization.
 In molecular dynamics simulations the two subsystems 
 are separated with 
 the elastic scattering assisted flexible 
 inner region [Kirchhoff et. al. JCTC, 2021, 17, 9, 5863] 
 -- ensuring a smooth transition in the radial distribution at the boundary between the two subsystems.
 %
\end{abstract}

\maketitle

\section{Introduction}

Modelling electrocatalytic reactions at the interface between solid 
and liquid requires capturing both the quantum-mechanical processes 
at the electrode, such as bond breaking and formation, and charge 
transfer, as well as the complex and fluctuating response of the 
surrounding electrolyte, which includes long-range electrostatic 
interaction with e.g., mobile ions, and the polarization response of 
the liquid. This presents a considerable challenge, and, therefore, 
different aspects of the electrocatalytic process are often modeled 
independently with established methodological frameworks~\cite{egill2017,gross2022ab,govindarajan2025} 
(for a recent review see~\cite{small2026}). 

In most cases the workhorse is Kohn-Sham density functional theory~\cite{hohenberg1964,kohn1965} (KS-DFT). Standard methods 
based on DFT, while powerful, must be extended or coupled with 
additional frameworks to account for the rich complexity of 
electrocatalytic systems. While ab initio molecular dynamics 
(AIMD) simulations have provided valuable insight into the fluctuation 
of the complex environment they are computationally demanding, 
particularly when the goal is to obtain a quantitative estimate 
of solvation energy, and the effect of applied voltage
and charge transfer at the electrode-electrolyte interface \cite{kirchhoff2023challenge,huang2023comparing,gross2023challenges}. 
AIMD simulations are typically limited to a timescale
of only a few picoseconds, whereas sampling over nanoseconds is 
often necessary to achieve proper equilibration and convergence 
of statistical averaging \cite{dawson2018equilibration,goswami2024solvation}.

To overcome these challenges, the hybrid quantum mechanics/molecular 
mechanics (QM/MM) \cite{warshel1976theoretical} strategy is a promising approach. 
In a QM/MM simulation, the system is partitioned into subsystems. 
The chemically active QM subsystem, where bond rearrangements and electronic 
effects take place, is treated using an electronic structure method. 
The remaining part is the MM subsystem and is described using a 
potential energy 
function depending only on atomic coordinates. This partitioning scheme 
dramatically reduces computational effort for large systems 
while retaining the accuracy needed to describe reactive events at the 
catalytic site. QM/MM can be a powerful tool for simulating 
electrocatalytic 
reactions in a realistic way.

QM/MM approaches are categorized by how the QM and MM 
subsystems are coupled: mechanical embedding (ME), 
electrostatic embedding (EE), and polarizable embedding (PE). 
In ME, there is no explicit QM/MM interaction, 
so long-range electrostatic effects are absent. 
In EE~\cite{field1990combined,dohn2020multiscale,dohn2017grid}, the 
MM subsystem is represented by fixed point charges, allowing 
the QM electronic density to respond to the electrostatic 
field of the MM region, partially capturing electrolyte 
effects on interfacial reactions. EE-QM/MM has been applied 
to estimate the free energy of water at the Pt(111)/water 
interface \cite{abidi2023electrostatically} and to study 
the hydrogen evolution reaction (HER) on MoS$_2$
electrodes \cite{clabaut2020solvation}, demonstrating its 
practicality for long-timescale dynamics otherwise inaccessible 
to AIMD. However, due to rapidly fluctuating near-surface 
electric fields and the electrical double layer extending 
well into the bulk liquid, non-mutually responsive EE-QM/MM 
models are inadequate for describing the solid/aqueous interface.

The PE-QM/MM approach overcomes the limitations of ME and 
EE by explicitly accounting for polarization 
response in both the QM and MM subsystems. 
The QM charge distribution induces 
dipoles, and possibly higher order moments, 
in the MM subsystem, and these in turn contribute 
to the 
Hamiltonian of the QM subsystem as an external potential, 
resulting in a self-consistent polarization field. 
Consequently, the energy of the system is a functional of 
the total polarization field. 
Sophisticated PE-QM/MM 
implementations including electrical dipoles have been developed
in the context of electronic excitations of solvated molecules \cite{jensen2003discrete,olsen2015polarizable,sneskov2011polarizable,lipparini2012linear,zeng2015analytic,loco:2016,list2016excited,loco2017hybrid,Menger2017}.
Higher order expansion up to the hexadecapole, which has been found 
to be adequate for typical intermolecular distances in water
\cite{batista00hexadeca}, have also been presented \cite{jonsson2019polarizable,dohn2019polarizable}.

While the PE-QM/MM approach provides a more rigorous treatment of 
interfacial polarization than ME or EE, 
it has so far not been applied extensively to electrocatalytic system. 
PE-QM/MM frameworks 
have only recently been extended to solid/liquid interfaces and their 
implementation in standard simulation packages is still maturing, 
partly due to the lack of transferability of polarizable force fields.
A step in the direction of electrocatalysis is a calculation of 
the Raman frequencies of some surface-bound intermediates in 
CO$_2$ electroreduction \cite{naserifar2021artificial}.

On another facet of hybrid simulations, machine learning interatomic potentials (MLIPs) have recently emerged as a powerful complement to both AIMD and classical QM/MM approaches for modeling solid/liquid interfaces, offering near-first-principles accuracy at a fraction of the computational cost and enabling access to nanosecond timescales and system sizes that are otherwise inaccessible \cite{HDNNP,H2OMLMD, MLSimple}. Architectures such as NequIP \cite{batzner2022}, MACE \cite{MACE2024}, and Allegro \cite{nomura2025} have been applied to metal/water interfaces, successfully reproducing structural and dynamical properties \cite{MLChallenge}.

In an ML/MM framework, the QM Hamiltonian is replaced by a neural network potential (NNP) embedded in a classical MM environment, dramatically accelerating free-energy sampling while retaining an accurate description of the reactive region \cite{MLMM}. Delta-learning strategies have proven particularly effective for condensed-phase simulations \cite{MLMMSemi}, and recent work has extended this paradigm to electrochemical systems by combining field-dependent MLIPs with ML electron-density response models to simulate metal–electrolyte interfaces under applied potentials. However, most MLIPs rely on a locality assumption that fundamentally limits their ability to capture long-range electrostatic interactions essential at charged interfaces, in polar solvents, and for charge-transfer reactions \cite{MLIPDielect, Feng2025}. While frameworks such as Latent Ewald Summation (LES) offer a route to incorporating long-range Coulomb interactions without requiring ambiguous DFT partial charge labels, their extension to heterogeneous interfaces with spatially varying dielectric permittivity remains an open challenge \cite{LES, Kim2025}, and proposed remedies for describing global charge redistribution at metallic electrodes under applied bias are still maturing \cite{LES}.

At the QM–MM boundary, the NNP inherits limitations from both the reference method and the embedding scheme, including inconsistent charge partitioning and the absence of mutual polarization under mechanical embedding \cite{Pultar2025}. Transferability remains a persistent concern, as ML models trained on specific electrode-electrolyte combinations frequently fail when extrapolating to unseen chemical environments or configurations outside the training distribution, producing unphysical energy drifts and trajectory instabilities \cite{NonTransferMLIP}. Collectively, these shortcomings — the locality assumption, inadequate treatment of long-range electrostatics, and lack of transferable polarizable MM force fields for the electrolyte — further motivate the development of PE-QM/MM frameworks.

Potential energy functions with high-level description of the electrostatics 
and thereby transferable to different environments have been developed for 
water and acetonitrile 
\cite{jonsson2019polarizable, dohn2019polarizable, naserifar2018quantum, EOJ2022, myneni2022polarizable}. These potential functions are typically parameterized by fitting results of quantum mechanical calculations (such as energy and atomic forces, multipole moments, and polarizabilities) of the solvent molecules at a specified level of theory.

In the following, we describe the polarizable embedding methodology, including a brief overview of the single center multipole expansion model (SCME) \cite{EOJ2022, myneni2022polarizable} and the QM/MM interface implemented in the grid-based projector augmented wave code GPAW \cite{GPAW2024}. To validate the PE-QM/MM approach, we first analyze 2D periodicity in energy and electrostatic potential by calculating the Coulomb potential of a graphene sheet with a water molecule, comparing the influence of the multipole expansion against a pure QM reference. We then examine the QM/MM interaction energy of 2D periodic ice layers using different grid expansions to assess the precision of the embedding scheme. Next, MD simulations are employed to investigate the QM/MM interaction in a dynamic setting, where an isotropic real-space damping function is introduced at the QM/MM boundary to prevent over-polarization. Gold-water MD simulations demonstrate that the damping value significantly impacts the PE-QM/MM interaction, confirming its necessity. Finally, a graphene-water MD simulation is performed to analyze the solvent distribution along the zz
z axis relative to a pure QM calculation.

\begin{figure}[!t]
\centering
\includegraphics[width=\linewidth]{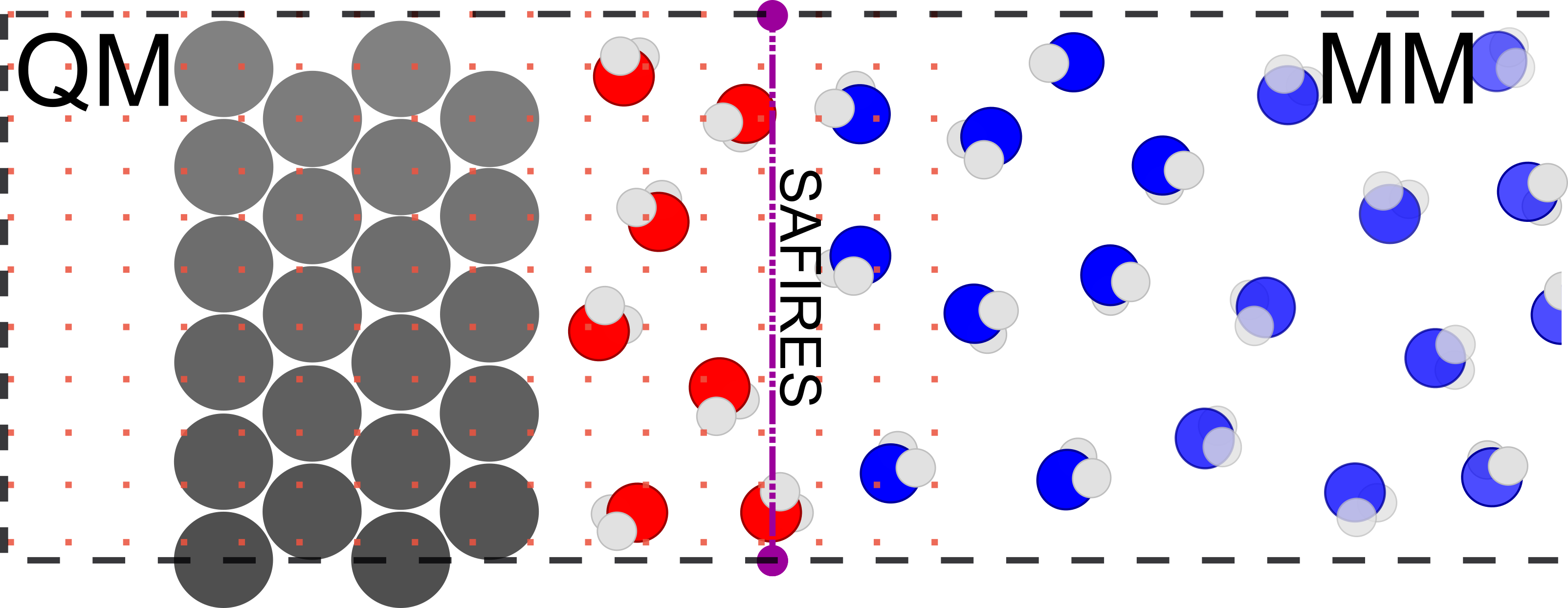}
\caption{Schematic of an example 2D periodic QM/MM interface. The QM region is composed of a slab (grey circles) and \ce{H2O} molecules  (red colored oxygen). The MM region is composed of \ce{H2O} molecules (blue colored oxygen). The orange squares represent the real space regular mesh grid points, $\bfr^g$, on which the QM system is described. Crossing of \ce{H2O} molecules between the two subsystems during e.g. MD simulations is prevented by the elastic {\it s}cattering {\it a}ssisted {\it f}lexible {\it i}nner {\it r}egion (SAFIRES \cite{kirchhoff2021elastic}) partitioning scheme. }
\label{fig:intro}
\end{figure}

\section{Theory}

In Kohn-Sham density functional theory~\cite{hohenberg1964,kohn1965} 
the energy of an electronic system is given by:
\begin{equation}
E^\mathrm{KS} = T_s + E_\mathrm{C}[\rho]  + E_{xc}[n]  + \int V_\mathrm{ext}[\rho]d\bfr 
\end{equation}
Here, $T_{s}$ is the kinetic energy of the non-interacting electrons 
whose total density corresponds to the ground-state density of the 
interacting electrons. $E_\mathrm{C}[\rho]$ is the 
Coulomb energy of the charge density, $\rho$. 
$E_{xc}[n]$ is the exchange-correlation energy functional
of the electron density, $n$, 
a functional form which is not known exactly and must be approximated 
in practice. It is typical in KS–DFT studies of electrocatalytic systems
that a local or semilocal approximation of the exchange-correlation energy is used. This transforms 
the problem of N interacting electrons to a problem of three spatial 
variables to describe the total electron density. Therefore KS–DFT
provides a feasible way to simulate systems with thousands of electrons.
Finally, the $V_\mathrm{ext}$ term is an external potential and is include 
in QM/MM 
to couple the QM subsystem to the MM subsystem.

In QM/MM the total energy of the system is split into the energy of each subsystem 
(QM and MM) and the explicit interaction between them (QM/MM)
\begin{equation}
    E^\mathrm{tot} = E^\rQM + E^\rQMMM + E^\rMM .
\end{equation}
The functional derivative of the $E^\rQMMM$ term with respect to the 
QM charge density defines the external potential in the QM subsystem due to the MM multipoles,
and the functional derivative with respect to the MM multipoles defines the potential fields in the MM subsystem due to the QM charge density.
An iterative scheme can then be used to solve for the self-consistent polarization field in the total system. The total energy of the system is therefore expressed as a functional of both the QM charge density and multipole moments of the MM system. The resulting energy functional can be expressed as
\begin{align}
    E^{\mathrm{tot}}&[\rho,M'] = \nonumber \\ 
     &E^\rQM[\rho] + E^\rQMMM[\rho,M'] + E^\rMM[M']
     \label{eq:qmmmtot}
\end{align}
where $M'$ are the total moments (permanent plus polarized) of the MM system. The electrostatic coupling between the QM charge density and the MM environment can be expressed as an interaction energy functional,
\begin{equation}
    E^\rQMMM_\mathrm{elst.} = \int \rho(\mathbf{r})V^\mathrm{MM}(\mathbf{r})d\mathbf{r} = \sum_i^{N_\rMM}\sum_s(M^i_s + \Delta M^i_s)V^{i\rQM}_s
\end{equation}
where $\rho(\mathbf{r})$ is the charge density of the QM subsystem, $V^\mathrm{MM}(\mathbf{r})$ the electrostatic potential in the QM region due to the MM sites, including contributions from both permanent and polarized moments, $V_s^{i\mathrm{QM}}(\mathbf{r})$ is the potential field of rank $s$ due to the QM charge density evaluated at MM site $i$. $M^i_s$ are the permanent moments  and $\Delta M^i_s$ are polarized moments of rank $s$, for MM site $i$. The MM potential field can be written as a sum over MM sites $i$, which in Einstein notation is
\begin{equation}
    V^\mathrm{MM}(\mathbf{r}) = \sum_i^{N_\rMM}\sum_s T^{ri}_s(M^i_s + \Delta M^{i}_s)
    \label{eq:MMtoQM}
\end{equation}
where $T^{ri}_s$ is the Coulomb interaction tensor of rank $s$ (see Supporting Information S2) and $\bf{r}$ is a coordinate in the QM region (see FIG.~\ref{fig:intro}). The coordinate $\bf{r}$ and center of mass of MM site $i$ share a common origin, which defines the global reference frame of the total system. The on-site potential field, due to the QM charge density, of rank $s$ are given by  
\begin{equation}
    V_s^{i\rQM} = \int \rho(\mathbf{r})T_s^{ri} d\mathbf{r}
    \label{eq:QMtoMM}
\end{equation}
The $E^\rMM[M']$ potential function requires solving self-consistently for $\Delta M$, and is in turn similarly affected by the presence of the QM charge density via the $E^\rQMMM[\rho,M']$ term. The details of this coupling depends on the polarizable potential function.

\subsection{Single Center Multipole Expansion}

In the SCME model the energy is a functional of the polarized moments, $\Delta M^i_s$, 
ascribed to the \ecom\ of \ce{H2O} molecule $i$ and is given by
\begin{equation}
    E^\mr{MM}[\{\Delta M^i_s\}] = E_\mr{elst}[\{\Delta M^i_s\}] 
                      + E_\mr{non-elst} \label{eq:SCME_tot}
\end{equation}
where the terms on the right hand side are, $E_\mr{elst}[\{\Delta M^i_s\}]$, 
the total electrostatic energy functional and
the non-electrostatic terms, $E_\mr{non-elst}$, which includes terms such 
as a pair-wise repulsive and short to intermediate range potential functions. 
In the case of the flexible variant, SCME/f, a $E_\mr{mon}$ term is also included 
which is a sum of the internal energies described by the Partridge-Schwenke~\cite{Partridge1997TheDO} 
potential energy surface (PS--PES) of the water monomer.

The electrostatic term can be split up into three contributions
\begin{align}
 E_\mr{elst}[&\{\Delta M^i_s\}] \nonumber \\
 &= E_\mathrm{perm}[\{\Delta M^i_s\}] + E_\mathrm{pol}[\{\Delta M^i_s\}] + E_\mathrm{self}[\{\Delta M^i_s\}]
\end{align}
accounting for the inter- and intramolecular contribution to the total electrostatic energy in the system. In the SCME model each MM site $i$ is assigned a permanent dipole up to, and including, a hexadecapole ($M^i=\{\mi_\al,\qi_\ab,\oi_\aby,\hi_\abyd\}$), as well as a dipole-dipole, dipole-quadrupole, and quadrupole-quadrupole polarizability (resulting in $\Delta M^i=\{\dmi_\al,\dqi_\ab\}$). At self-consistency the first two terms combine to give
\begin{align}
 E_\mr{perm+pol}[\{\Delta M^i_s\}] = \frac{1}{2}&\sum_i^{N_\mathrm{MM}}(\mu^i_\al+\Delta\mi_\al)V^i_\al  \nonumber \\
 &\ \ + (\theta^i_\ab + \Delta \qi_\ab)V^i_\ab \nonumber \\
 &\ \ + \Omega^i_\aby V^i_\aby + \Phi^i_\abyd V^i_\abyd
\end{align}
$E_\mr{self}$ is the on-site self-energy, given by 
\begin{equation}
    E_\mr{self}[\{\Delta M^i_\al\}] = - \frac{1}{2}\sum_i^{N_\mathrm{MM}}\left(\Delta\mi_\al V^i_\al 
  + \frac{1}{3}\Delta\qi_\ab V^i_\ab\right)
\end{equation}
and accounts for the cost in energy to polarize the molecules. In the SCME model, the Coulomb interaction tensor is damped using a Gaussian-based scheme~\cite{Stone:2011}, such that the rank-0 damped Coulomb interaction tensor becomes
\begin{equation}
 T^{ij,d}_0 = \frac{1}{r^{ij}}\lambda_0(r^{ij})
\end{equation}
where $\lambda_s(r_{ij})$ is a rank-$s$ damping function which depends on the distance between the center of mass of the $i$ and $j$ MM sites (see Supplementary Information S2). The potential at site $i$ is given by
\begin{equation}
 V^i = \sum_{j\neq i}^{N_\mathrm{MM}}\sum_s T^{ij,d}_s(M_s+\Delta M_s)
 \label{eq:MMtoMM}
\end{equation}
and, in the case of SCME with dipole-dipole, $\al$, dipole-quadrupole, $A$, and quadrupole-quadrupole polarizabilities, $C$, the induced moments are
\begin{align}
 \dmi_\al =& \al^i_\ab V^i_\be + A^i_{\al,\by} V^i_\by \\
 \dqi_\ab =& A^i_{\gamma,\ab}V^i_\gamma + C_{\ab,\gamma\delta}V^i_{\gamma\delta}
 \label{eq:indmom}
\end{align}
This set of linearly coupled equations, Eqs.~\eqref{eq:MMtoMM}-~\eqref{eq:indmom}, is converged iteratively.

\subsection{PE-QM/MM Energy Functional}
The total energy functional expression in Eq.~\eqref{eq:qmmmtot}
includes terms which define the interaction between the QM charge density with the permanent- and the polarizable-moments, and similarly define the interaction of the MM polarizable moments with the QM charge density. The coupling between the two subsystems become
\begin{align}
 \frac{\partial E^\rQMMM}{\partial \rho(\bfr)} =& V^\rMM(\bfr) \\
 \frac{\partial E^\rQMMM}{\partial (M^i_s+\Delta M^i_s)} =& V^{i\rQM}_s 
\end{align}
Therefore, in an iterative loop where the density and polarizable moments are updated self-consistently, the SCF equations are
\begin{equation}
 \Delta M^i_s = \sum_ta_{s,t}(V^{i\rMM}_t + V^{i\rQM}_t)
 \label{eq:totPol}
\end{equation}
where the on-site potential field of rank $s$ at MM site $i$ are given by eqs.~\ref{eq:QMtoMM} and~\ref{eq:MMtoMM}, and $a_{s,t}$ is a general polarizability which expresses the linear response of moment of rank $s$ to potential field of rank $t$. This defines the total potential field at MM site $i$, or
\begin{equation}
 V^{i\mathrm{Tot}}_s = V^{i\rMM}_s  + V^{i\rQM}_s 
 \label{eq:totMM}
\end{equation}
Using this we can write the total permanent plus polarizable interaction potential function as
\begin{align}
 E^\rQMMM_{\mathrm{pol}+\mathrm{perm}} +& E^\rMM_{\mathrm{pol}+\mathrm{perm}} = \frac{1}{2}\int\rho(\bfr)V^{\rMM}(\bfr)d\bfr \nonumber \\
 +& \frac{1}{2}\sum_i^{N_\rMM}\sum_s(M^i_s + \Delta M^i_s)V^{i\mathrm{Tot}}
\end{align}
and self-energy
\begin{equation}
 E_\rself = -\frac{1}{2}\sum_i^{\rMM}\sum_s\Delta M^i_sV^{i\mathrm{Tot}}_s
\end{equation}

\section{Implementation}

\subsection{Finite-Difference Real-Space Grid Implementation}

PE-QM/MM energy functional is implemented in the open source grid-based projector augmented wave code GPAW~\cite{GPAW1,GPAW2,GPAW2024}.  The projector augmented wave (PAW) method~\cite{paw1,paw2,ivanov2025upaw} is used to describe the electrons near the nuclei, and the core electrons for each atom are frozen to the result of a reference scalar relativistic calculation of the isolated atom. 

In the real-space grid representation the wavefunctions describing the valence electrons 
are represented on a regular mesh of real-space grid points, $\psi(\br^G)$, and the valence electron density is represented on a finer mesh, $n(\br^g)$. The effective mean-field Hamiltonian is first evaluate using the density on the fine mesh and then transformed to the regular mesh 
$H^\mr{KS}_\mr{eff}(\br^g)\rightarrow H^\mr{KS}_\mr{eff}(\br^G)$ - therefore the MM potential in Eq.~\ref{eq:MMtoQM} is evaluated on the fine-mesh grid, and similarly the potential field on each MM site $i$ Eq.~\ref{eq:QMtoMM} is evaluated by integrating the charge density times tensor operator on the fine-mesh grid. 
The general expression for the Hamiltonian including the external potential due to the MM sites, as well as the resulting QM nuclei forces, in GPAW, can be found in the Supporting Information S1.

The real-space mesh representation of the Hamiltonian operator allows for general customization, such as mixed open and periodic boundary conditions, and hence is ideal for surface-liquid interfaces where the total system is $2D$ periodic. The mesh, which is used to describe the QM region, can therefore be kept minimal in size along the non-periodic axis, and only the periodic axes need to be matched between the two subsystems. In the following sections, $2D$ periodicity is assumed - and the non-periodic axis is the global $z$-axis. See FIG.~\ref{fig:intro}.

\subsubsection{SCF}

A self-consistent solution is reached for the total system with a dual SCF cycle \cite{jonsson2019polarizable,dohn2019polarizable} - and outer and inner loop. In the outer loop the QM charge density is updated while the polarizable moments of the MM subsystem are kept fixed. In an inner loop the polarizable moments (Eq.~\eqref{eq:totPol}) are solved for self-consistently including the potential fields from the QM subsystem - while the charge density fixed. Given a reasonable initial guess for 
the QM charge density, the total on-site potential Eq.~\eqref{eq:totMM} is 
calculated at each MM site $i$, and the polarizable moments evaluated
\begin{align}
 \Delta M_s =& \sum_t{a_{s,t}V^{i\mathrm{Tot}}_t} \nonumber \\
 =& \sum_t{a_{s,t}}\left(V^{i\rMM}_t + V^{i\rQM}_t \right) \label{eq:totPolMM}
\end{align}
Eqs.~\eqref{eq:totPolMM} and \eqref{eq:MMtoMM} are iterated until a  
convergence threshold is reached. 
The convergence threshold for the polarizable moments is tied with the residual 
difference of the electron density of the QM region -- it is set to 
$c_{pol}=\mathrm{min}[abs(\Delta n),10^{-3}]$, i.e. 
it is updated and based on the absolute residual difference 
($abs(\Delta n)$) of the QM electron density between QM-SCF steps.

In some cases it can be beneficial to include a polarization mixing 
term (similar to density mixing in implementations of KS-DFT), such that
\begin{equation}
 \Delta M^i_{s,\mathrm{new}} = (1 - t)\Delta M^{i}_s  + t\Delta M^i_{s,\mathrm{prev}}
 \label{eq:MMPolMixing}
\end{equation}
which has proven to stabilize the SCF cycle in symmetric arrangements of highly polarizable SCME centers~\cite{myneni2022polarizable}, and can similarly be applied in QM/MM simulations.

\subsection{Tensor Damping Functions}

The position of MM molecules in the global reference frame will possibly place them within the QM grid space as illustrated in FIG.~\ref{fig:intro}, and hence close to, or on top, of a grid point. 
Potential fields in terms of the interaction tensors can therefore
diverge, resulting in what is commonly known as the polarization catastrophe, as coined by Thole~\cite{thole:1981}. In order to avoid this catastrophe, the tensor damping functions are smeared out, and the point moments are described with  
a screened interaction that captures the effect of the overlap of charge densities.

Common choices include Thole type~\cite{thole:1981} damping 
functions~\cite{masia2005,masia2006,burnham1999} which are based on an exponential decay description of point charges or a normalized Gaussian description~\cite{Stone:2011}. A more 
comprehensive comparison of damping functions is beyond the 
scope of this work, but can be found 
elsewhere.~\cite{dampingbothyeah,masia2005,masia2006} 
Gaussian type damping functions are used in this work due to their simple recursive relation (see Supplementary Information S2).

\begin{figure}[!t]
\centering
\includegraphics[width=\linewidth]{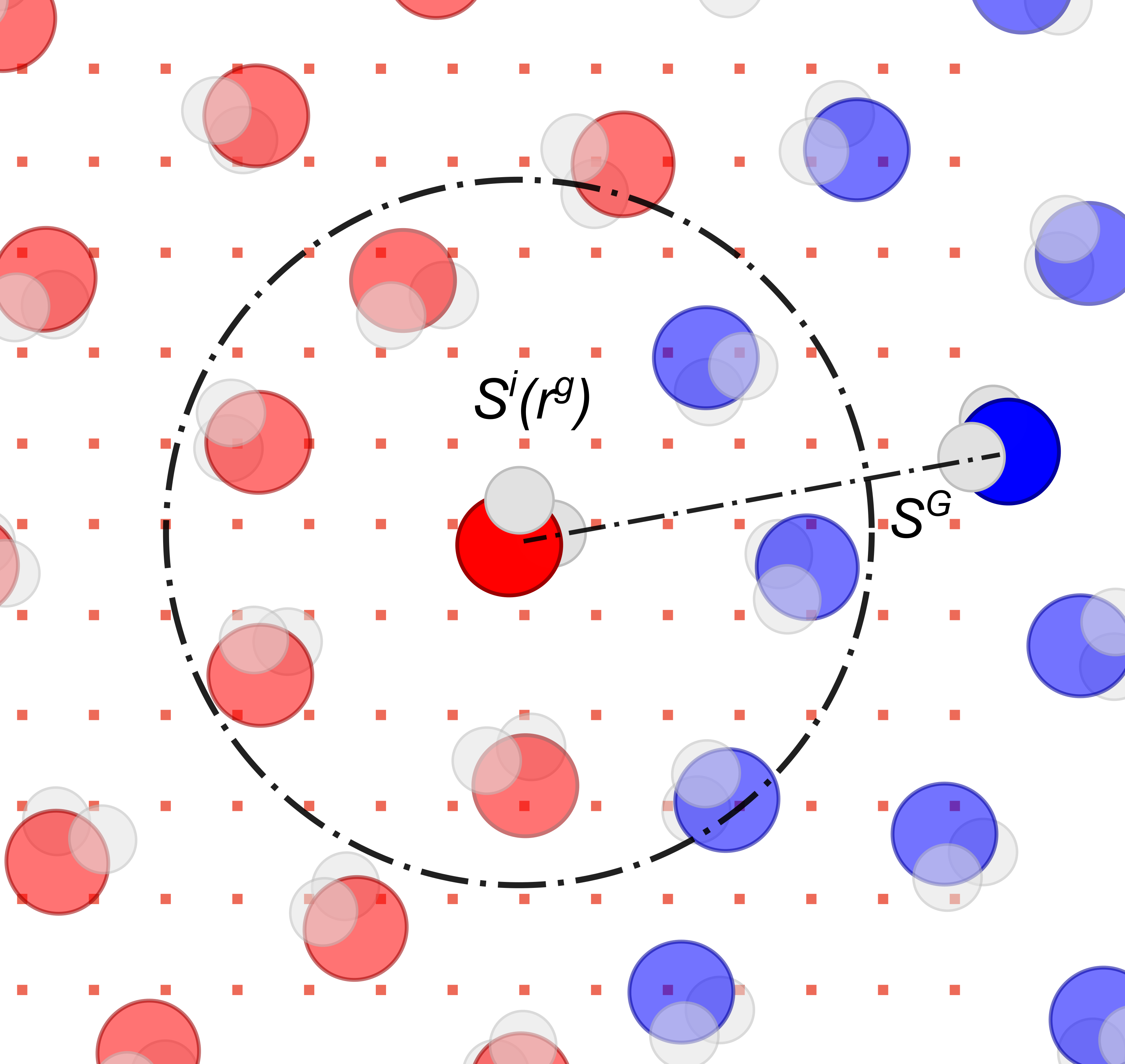}
\caption{Schematic of the isotropic damping between an QM and MM \ce{H2O} molecule. The distance between the centers of mass determines the damping factor $S^G$, and all grid-points (orange) in the circular area are damped according to $S^i(\bfr^g)$, Eq. \ref{eq:isodamp}. The damping the region around the QM \ce{H2O} molecule does not depend on the angle or rotation of either molecule, only on the radial distance, it is therefore isotropic.}
\label{fig:isodamp}
\end{figure}

\subsubsection{Isotropic Real-Space Damping}

A key issue in previous PE-QM/MM implementations \cite{jonsson2019polarizable,dohn2019polarizable} is that the electrostatic damping between the QM and MM subsystems is anisotropic, as it is based on the distance between the center of mass of each MM molecule and the real-space grid points $\br^g$. This leads to inconsistent damping depending on molecular orientation: for example, damping is strong when the MM \ce{H2O} molecule is the hydrogen donor in a \ce{H2O} dimer, due to significant electron density overlap from the lone pair of the QM oxygen, but is nearly absent when the QM molecule is the hydrogen donor instead. In many-body systems, strong local electric fields can arise between \ce{H2O} pairs even at non-optimal angles through many-body polarization propagating throughout the system. Since only the QM molecule carries an explicit electron cloud, the electrostatic interaction at such orientations may go entirely undamped, posing a risk of over-polarization.

An alternative way to introduce damping is to enforce a dependence on the distance between pairs of QM and MM \ce{H2O} molecules -- that is a radial dependence and therefore independent on the relative orientation of the pair.
To achieve this, we rewrite the QM/MM electrostatic energy as
\begin{equation}
    E^{\rQMMM} = \sum_{i}^{N_\rMM}\int S^i(\bfr)\rho(\bfr)V^i(\bfr)d\bfr
\end{equation}
where we define a spatially resolved damping function, $S^i$, as
\begin{equation}
    S^i(\bfr) = \sum_{\substack{a\in\rQM \\ |\bfr^a-\bfr^i|\leq R_{c,\mathrm{d}}}}S^{\mathrm{G}}(|\bfr^a - \bfr^i|)w^a(\bfr^a,\bfr)
    \label{eq:isodamp}
\end{equation}
Finally, the center of mass distance dependent damping factor is given by
\begin{align}\label{eq:S_G_beta}
    S^{\mathrm{G}}(|\mathbf{r}^a-\mathbf{r}^i|)
    &= \mathrm{erf}\!\left(\beta|\mathbf{r}^a -\mathbf{r}^i|\right) \nonumber\\
    &\quad - \frac{2}{\sqrt{\pi}}\,\beta|\mathbf{r}^a -\mathbf{r}^i|
      \exp\!\big(-\beta|\mathbf{r}^a -\mathbf{r}^i|^2\big)
\end{align}
where $0 \leq S^{\mathrm{G}}(|\bfr^a - \bfr^i|) \leq 1$, and $\beta \in \Re^+_{0}$.
The index $a$ indexes the centers of mass of QM \ce{H2O} molecules, and $R_c$ is a cutoff radius. See FIG.~\ref{fig:isodamp}. Therefore, for a given MM \ce{H2O} molecule $i$, one needs to consider a subset of neighboring QM \ce{H2O} molecules when constructing the regional damping function.  
$w^a(\br^a,\br)$ are appropriate weight functions which partitions the real-space fine mesh into equally weighted regions around each QM \ce{H2O} center of mass. See Appendix~\ref{app:isotropic}. 

\begin{figure}[!t]
\centering
\includegraphics[width=\linewidth]{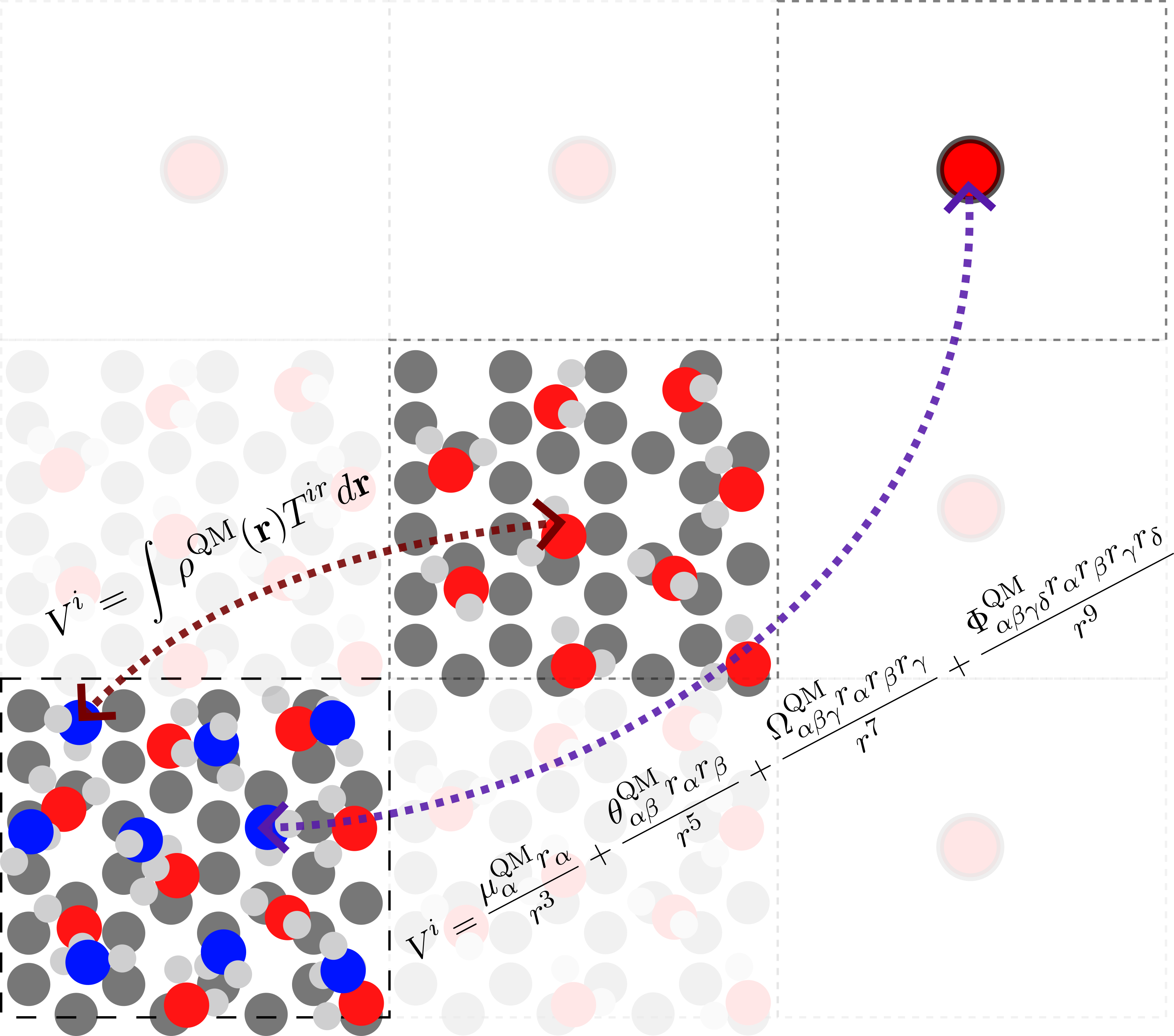}
\caption{Schematic of the multipolar expansion approximation between the QM and MM systems. In the near-neighbour, the interaction is calculated as explicit integrals of the QM charge density and the appropriate tensor operator to generate fields and higher order gradients on each MM site $i$. In the outer region, the QM charge density is instead represented as a single multipole expansion point, and the potential field at each MM site is calculated via. the same tensor expansion as in SCME.}
\label{fig:expansion}
\end{figure}

\subsection{Embedded Electrostatic Interaction}

The electrostatic interaction between the two subsystems is divided into a near- and far-field term using a lattice translation operator
\begin{equation}
 \mathcal{P}(\mathcal{N},\br^i,\br^j) = (\br^i - \br^j) + \mathbf{R}_c \cdot \mathcal{N} 
\end{equation}
where $\mathbf{R}_c$ is the lattice vector of the total systems cell, and $\mathcal{N}$ are positive and negative integer multipliers describing lattice translations. 
The potential at MM site $i$ is given by
\begin{align}
 V_s^{i\rQM} =& \sum^{\mathcal{N}}_{(0,0,0)}\int T^{ir}_s[\mathcal{P}(\mathcal{N},\mathbf{R}^i, \br)]\rho(\br)d\br \nonumber \\
 +& \sum_{(n_x,n_y,0)}^{\mathcal{N}'}\sum_s T_s^{ir}[\mathcal{P}(\mathcal{N}',\mathbf{R}^i,\br)]M^{\rQM}_s
\end{align}
where $\mathcal{N}$ and $\mathcal{N}'$ are vectors of integers outlining the near-neighbor and outer-neighbor periodic images.
For outer-neighbor images the potential field at MM site $i$ is reduced to an interaction with a single expansion point describing the QM charge density, where
$M_s^{\rQM}$ is the rank $s$ multipole moment evaluated for the QM system. See FIG.~\ref{fig:expansion}. 
The QM Cartesian multipole expansion center is chosen to be at the center of global system with respect to the periodic axis $x,y$ whereas the $z$ coordinate is chosen to be at the center of mass $z$ coordinate of the QM H$_2$O molecules. 

Similarly, the external potential in the QM region is evaluated as
\begin{align}
 V^{\rMM}(\bfr) &= \sum^{\mathcal{N}}_{(0,0,0)}\sum_i^{N_\rMM}\sum_s T^{ri}_s[\mathcal{P}(\mathcal{N}',\mathbf{R}^i,\br)](M^i_s + \Delta M^{i}_s) \nonumber \\
 +& \sum_{(n_x,n_y,0)}^{\mathcal{N}'}\sum_j^{C_\rMM}\sum_s T_s^{ir}[\mathcal{P}(\mathcal{N}',\mathbf{R}^i,\br)]M^{C}_s
    \label{eq:MMtoQM2}
\end{align}
that is, for outer-neighbor images, the number of MM sites are reduced to $C_\rMM$ centers using a clustering and origin shift algorithm, as described in the next section, leading to a greatly reduced set of explicit numerical evaluations on the real-space fine grid mesh. In the remaining sections the inner $\mathcal{N}$ and outer $\mathcal{N}'$ expansion grids are defined by $\mathcal{N}:=Nc$ ({\it n}umber of inner {\it c}ells) and 
$\mathcal{N}':=Nc_{outer}$ (number of {\it outer} cells).

\begin{figure}[!ht]
\centering
\includegraphics[width=\linewidth]{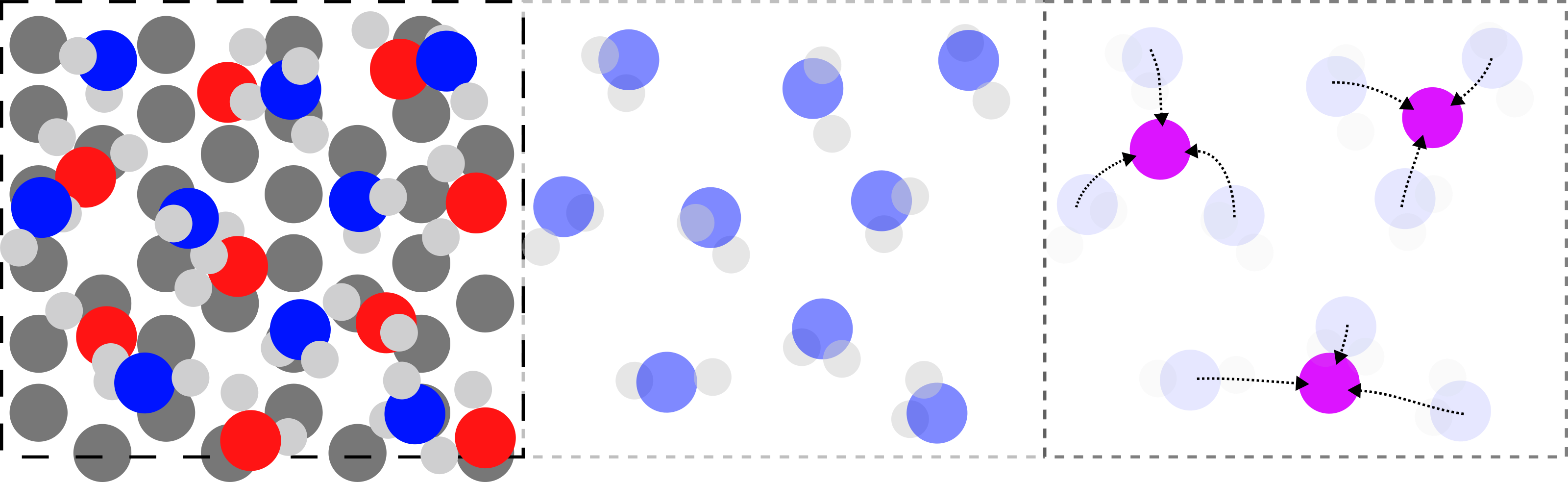}
\caption{Origin-shift algorithm for compressing distributed MM multipole contributions onto a reduced set of expansion centers. (Left) The QM region with the MM overlay region containing many individual \ce{H2O} molecules(gray, blue, and red spheres). (Center) The first nearest neighbour cell of \ce{H2O} molecules where no clustering is considered. (Right) Application of a K-means clustering algorithm that partitions the MM sites into spatially coherent groups, with each cluster represented by an effective expansion center (cluster center) (larger magenta/pink spheres). This results in origin-shifted multipole moments which have been translated from each site's local origin to a cluster center.}
\label{fig:sklearn}
\end{figure}

\subsection{Far-Field Electrostatic Interactions}
In the long-range regime of the electrostatic interaction, we perform a Cartesian multipole expansion of the QM charge density. On the MM side, rather than retaining every individual site moment, we accumulate and origin-shift the distributed MM multipole contributions onto a reduced set of expansion centers. These centers are determined from the spatial distribution of MM sites using a K-means clustering algorithm \cite{kmeans}, so that the entire outer MM region is represented by a small number of effective multipole points. This compression drastically reduces the number of interacting terms without sacrificing accuracy in the long-range limit. FIG.~\ref{fig:expansion} presents an example interaction between an MM site $i$ and the periodic replica of the QM subsystem. The clustering algorithm implemented allows for a variable number of origin-shifted sites, see FIG.~\ref{fig:sklearn}. See the Supplementary Information S5 for the K-means algorithm used and for the expressions used to shift the multipole moments (dipole to hexadecapole) from the center of mass of the molecule (where it is defined, and what we refer to as the 'origin' in space for the multipole moment) to a new common center, which in this case are the cluster centers.

\section{Computational Details}


In all calculations, the base exchange-correlation energy functional is PBE~\cite{PBE}. Wavefunctions are described on a real-space grid with finite-difference methods, and the grid spacing is varied and stated in the figure captions. 
Only the Gamma-point was used to sample the Brillouin Zone (BZ) for all calculations until the density, eigenstates, and self-consistent total electronic energy convergence thresholds reached $10^{-8}$~e, $4\times 10^{-8}$~eV$^2$, and $5\times 10^{-4}$~eV, respectively. 
Additionally, for the PE-QM/MM calculations, we set the convergence criterion for the maximum absolute difference between the change in the MM dipole and the MM quadrupole to $10^{-8}$~a.u., and the damping value $\beta$ from the isotropic damping factor in equation \eqref{eq:S_G_beta} to 0.291 \AA$^{-1}$. The cutoff for the damping was set to 5.90\ \AA\ to match the water solvation shell.
In the case of the gold-water QM/MM MD simulations, the maximum absolute difference between the change in the MM dipole and the MM quadrupole was changed to $10^{-6}$~a.u., and a inner grid $Nc$ = [1,1,0] with a outer grid expansion of $Nc_{\mathrm{outer}}$ = [9,9,0] was applied. 
The gold-water system contains 48 gold atoms and 32 \ce{H2O} molecules, while 12 \ce{H2O} molecules belong to the QM region and 20 \ce{H2O} molecules belong to the MM region. The two subsystems are separated using SAFIRES.\cite{kirchhoff2021elastic}
The cell parameters are set to 8.74\ \AA, 10.09\ \AA, and 30.42\ \AA\ with a vacuum region of 6.65\ \AA\ placed below the gold slab.
The graphene-water QM/MM MD simulation was performed using $10^{-6}$~e, and $10^{-6}$~eV$^2$, as convergence criteria for the density and eigenstates, respectively. Furthermore, the maximum absolute difference between the change in the MM dipole and the MM quadrupole was kept at $10^{-6}$~a.u., with a real-space grid of $h$ = 0.20 and a inner grid of $Nc$ = [1,1,0] with a outer grid expansion of $Nc_{\mathrm{outer}}$ = [9,9,0]) was applied. The simulation cell contains a fixed carbon sheet consisting of 32 carbon atoms, and 32 \ce{H2O} molecules, whereas 8 molecules belong to the QM region and 24 molecules belong to the MM region. The cell parameters are set to 9.80\ \AA, 8.49\ \AA\ and 26.34\ \AA\ with a vacuum region of 5.50\ \AA, placed below the graphene sheet and above the outermost QM \ce{H2O}. 


\subsection{Periodic Electrostatic Interaction Analysis}
\begin{figure}[t]
\centering
\includegraphics[width=\linewidth]{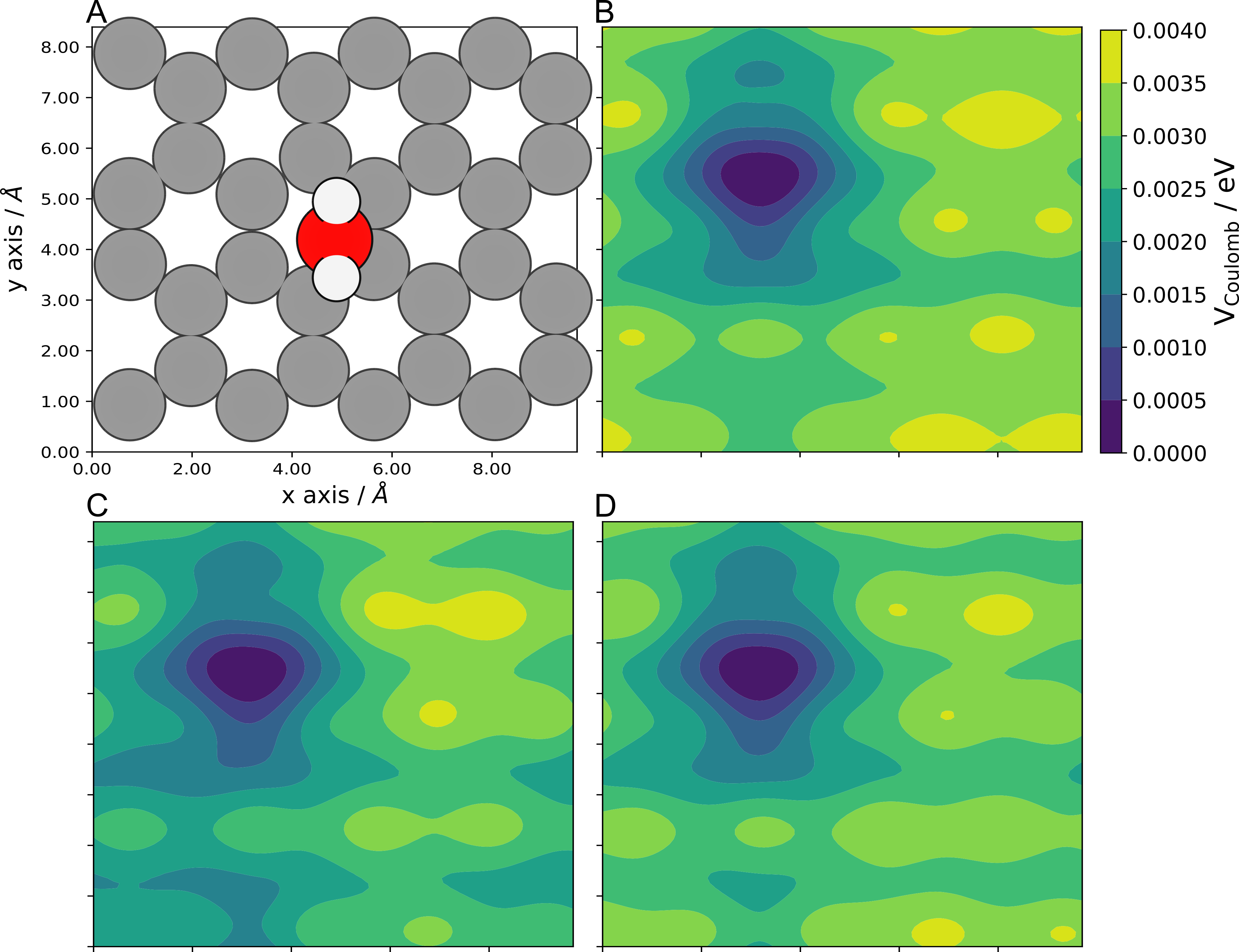}
\caption{Coulomb potential of a water molecule on top of a graphene layer. The position of the water molecule is shown in panel A, and the pure QM potential in the graphene layer is shown in panel B. Below, panels C and D provide the response potential of the PE-QM/MM calculation with a $Nc$ = [1,1,0] and $Nc_{\mathrm{outer}}$ = [0,0,0] and $Nc_{\mathrm{outer}}$ = [9,9,0], respectively. The QM/MM calculation was performed with a real-space grid of $h$ = 0.20.}
\label{fig:QMMM_pot_water_graphene}
\end{figure}

In FIG.~\ref{fig:QMMM_pot_water_graphene}, the Coulomb potential induced in the graphene sheet by a water molecule, evaluated in the two-dimensional unit cell, is compared for three levels of theory. In the upper-left panel, the top view of the geometric simulation setup is illustrated. In the PE-QM/MM calculation, the graphene sheet is assigned to the QM region, and the water molecule belongs to the MM region. 
Panel B shows the Coulomb potential in the graphene sheet calculated by the pure QM level of theory. In panels C and D, the PE-QM/MM approach is applied using different expansion configurations, while panel C shows the inner cell expansion of $Nc$ = [1,1,0] ($Nc_{\mathrm{outer}}$ = [0,0,0]), panel D shows the more accurate inner and outer cell expansion setting ($Nc$ = [1,1,0] and $Nc_{\mathrm{outer}}$ = [9,9,0]). 
All calculations provide a Coulomb potential minimum centered at ca. 3.0~\AA along the x-axis and 5.5 along the y-axis, reflecting the long-range polarization response of the graphene sheet to the water dipole moment. 
Notably, the diagrams in panels B and D exhibit a comparable 2D periodicity along both axes. Only the Coulomb potential in the PE-QM/MM calculation with the configuration $Nc$ = [1,1,0] and $Nc_{\mathrm{outer}}$ = [0,0,0] is inaccurately described at the boundaries. 
The Coulomb potential lacks the correct long-range modulation, underscoring the well-known sensitivity of 2D periodic systems to the number of image cells included in the lattice sum \cite{HOLZMANN_2005, Tyagi_2004}. 
Hence, introducing the outer cell expansion can significantly improve the electrostatic contribution with regard to the 2D periodicity. The isosurfaces in the more accurate QM/MM calculation in panel D and the pure QM calculation are in good agreement. The PE-QM/MM simulation captures mutual polarization between the water molecule and the graphene sheet and converges to a topology that closely reproduces the pure QM reference. The position and the depth of the central minimum, as well as the peripheral features near the cell corners, are reproduced accurately. 
The inclusion of long-range contributions is critical for convergence toward the full QM treatment in PE calculations. This benchmark therefore demonstrates that the 2D translational periodicity of the graphene substrate can be faithfully described within the PE-QM/MM framework, provided a sufficient number of near-neighbor image cells is included in the embedding potential.

\subsection{Periodic Ice-Ih Lattices}
\begin{figure*}[t]
\centering
\includegraphics[width=\textwidth]{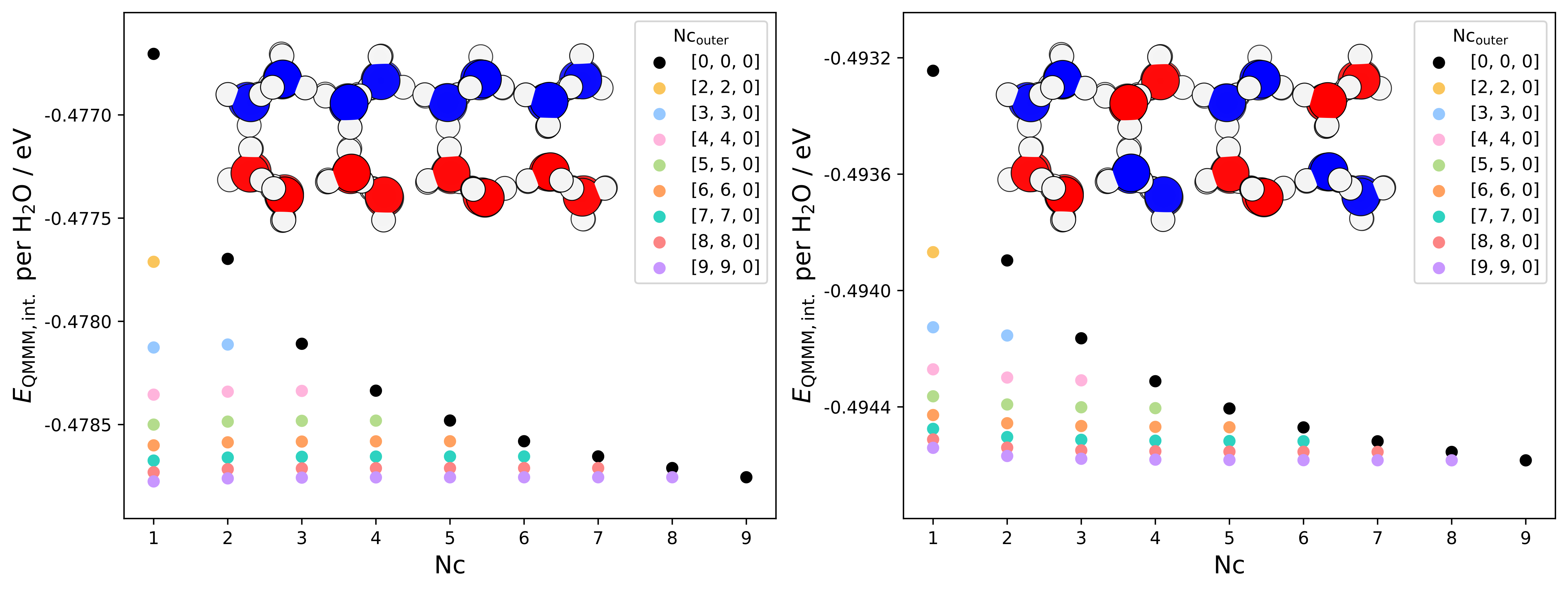}
\caption{QM/MM interaction energy of two ice layers containing QM (red) and MM (blue) molecules in a 2D periodic system calculated in different $Nc/Nc_{\mathrm{outer}}$ expansion configurations. In the left diagram, the QM/MM partitioning is layered, whereas in the right panel, it is mixed. The QM/MM interaction energy is normalized per water molecule, see Supplementary Information S4 for the equations. The QM/MM interaction energy is shown using the inner cell grid expansion. The black dots present the QM/MM calculation using only the inner cell grid expansion. In addition, the outer cell grid expansion is displayed in different colors. The QM/MM calculation was performed with a real-space grid of $h$ = 0.20.}
\label{fig:E_intvsNc_h20}
\end{figure*}
FIG. \ref{fig:E_intvsNc_h20} shows the QM/MM interaction energy per water molecule between the QM and MM region with respect to the number of copies of the original cell on the inner cell grid $Nc$. The blue molecules illustrate the QM sites, while the red molecules define the MM sites. 
The black dots present the calculations performed purely on the inner cell grid $Nc$ without an expansion on the outer cell grid ($Nc_{\mathrm{outer}}$ = [0,0,0]). Thus, the calculation with 9 copies of the original cell on the inner cell grid exhibits the most accurate QM/MM calculation. 
Introducing the expansion on the outer grid benefits the convergence pattern as the interaction energy values move closer to the convergence value at a lower inner grid expansion. 
The greater the expansion of the outer grid, the fewer inner grid copies are necessary to reach the convergence value. Hence, the computational effort can be reduced by setting $Nc$ = [1,1,0] and using a high value for the outer expansion of the grid, such as $Nc_{\mathrm{outer}}$ = [8,8,0] or $Nc_{\mathrm{outer}}$ = [9,9,0]. 
Moreover, the convergence pattern is independent of the QM/MM partitioning. Mixing the QM and MM sites by creating alternating rows of QM and MM sites provides a similar convergence pattern as in the layered system. In the mixed system on the right-hand side, the accuracy of a calculation with $Nc$ = [9,9,0] can already be achieved with $Nc$ = [1,1,0] and $Nc_{\mathrm{outer}}$ = [9,9,0].
In particular, the expansion of the outer grid not only allows an accurate description of the system, but also a significant speed advantage by more than 5 times in the case of $Nc$ = [1,1,0] with $Nc_{\mathrm{outer}}$ = [8,8,0] or $Nc_{\mathrm{outer}}$ = [9,9,0] compared to $Nc$ = [9,9,0], as shown in FIG.~S1 in the Supplementary Information. 
Beyond that, we examined the QM/MM interaction energy of the 2D periodic ice layers using different real-space grid spacing in the QM calculations. As shown in the SI, the convergence of the QM/MM interaction energy is independent of the real-space grid (FIG.~S2-S4).
Additionally, the QM/MM interaction energy values are in close alignment with the pure QM and pure MM calculations. As reported in tables S1 and S2 in the SI, the QM/MM interaction energy lies between the pure QM and pure MM interaction energy values for all investigated real-space grids.

\subsection{Gold-Water Interface}
\begin{figure}[t]
\centering
\includegraphics[width=\linewidth]{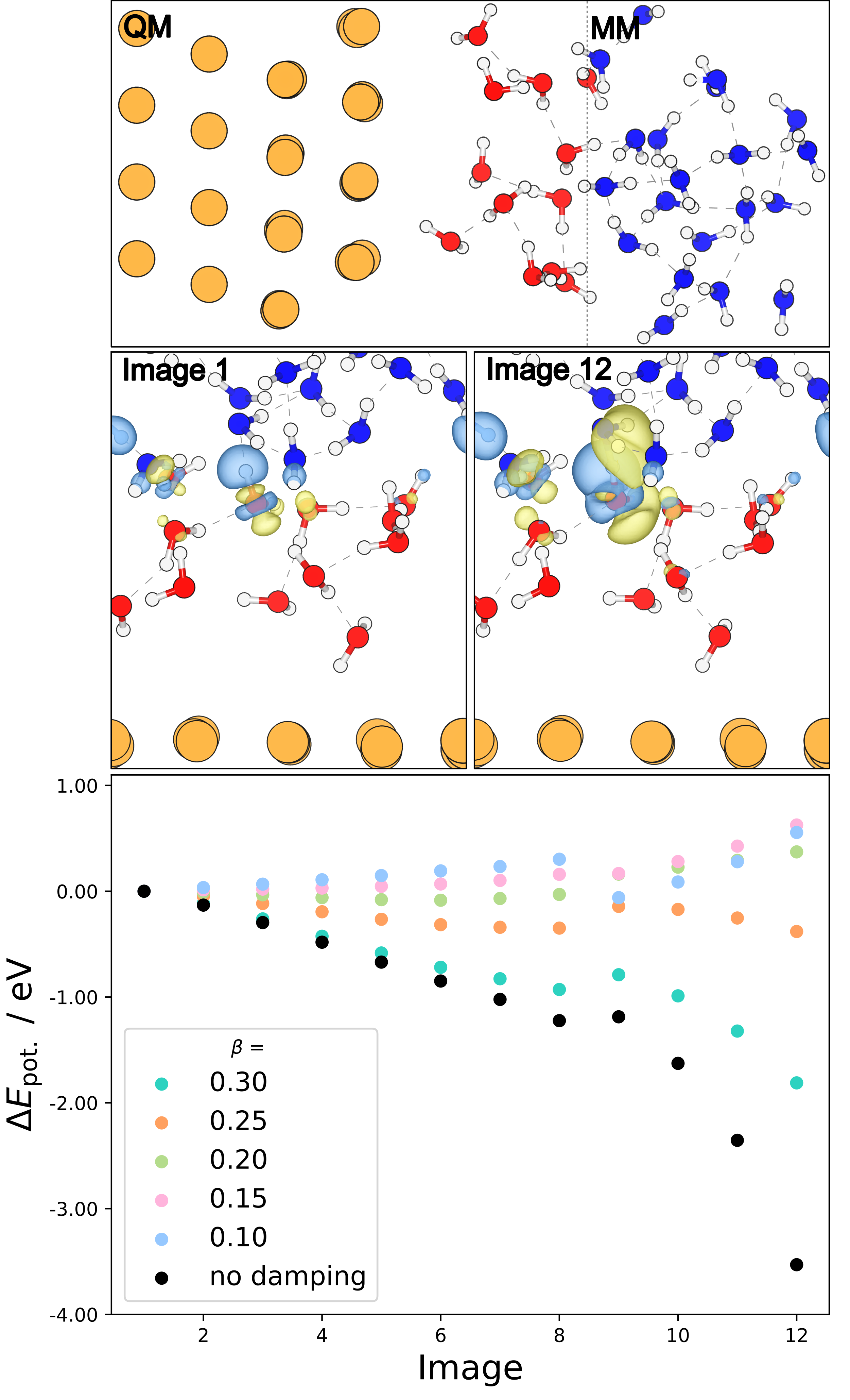}
\caption{See main text for the description of the top and middle panel. Bottom panel: potential energy of a gold-water PE-QM/MM MD simulation 12 steps before the polarization catastrophe. The diagram shows the potential energy of the PE-QM/MM simulation with and without isotropic damping. Above the diagram, the charge density difference between the damped and non-damped systems in the first and last images corresponding to the diagram is illustrated. The simulation was run with a real-space grid of $h$ = 0.25 in the QM region, and an inner cell grid of $Nc$ = [1,1,0] combined with an outer cell grid of $Nc_{\mathrm{outer}}$ = [9,9,0], for the QM/MM embedding. The isosurfaces were generated by VESTA with an isosurface level of $10^{-3}$\,$e$\AA$^{-3}$ \cite{VESTA}.}
\label{fig:MD_blow_up}
\end{figure}
FIG. \ref{fig:MD_blow_up} presents the potential energy comparison between the PE-QM/MM MD calculation with different isotropic damping values and without isotropic damping. Based on the energy diagram, the potential energy in the non-damped calculation reflects an exponential decline, leading to polarization catastrophe due to the dimer with the QM \ce{H2O} providing the hydrogen. At the same time, the potential energy of the calculations with the damped QM/MM interaction illustrates a rather constant energy profile, depending on the isotropic damping value. Decreasing the isotropic damping value $\beta$ to 0.20 or less provides an increase in the potential energy values. 
The course of potential energy values gives the impression that a value of $\beta \leq$ 0.20 is necessary.
However, taking into account the dimer binding curve from FIG.~S6 in the SI, the optimal value of $\beta$ lies between 0.27 and 0.32. In the case of $\beta$ = 0.291, the dimer binding curves of the QM/MM configuration fit between the pure QM and MM curves. Beyond that, FIG.~S7 reveals the scale of effective damping on the electrostatic interaction in the gold-water PE-QM/MM MD simulation by $S^G$. Each value of $\beta$ leads to a rather constant value of $S^G$ in the 12 images. FIG.~S7 indicates that $\beta$ = 0.30 dampens the electrostatic interaction already by 60~\%, and $\beta$ = 0.10 leads to a diminished electrostatic interaction of 5~\%. Moreover, $\beta$ = 0.10 would introduce a significant damping even when the QM and MM sites are 12~\AA\, apart, while for $\beta$ = 0.30 the effective damping fades at a distance of 7.50~\AA.
The difference in the charge densities visualizes the discrepancy between the damped calculation with $\beta$ = 0.291, and the non-damped calculation. In the first image, there is already a charge excess in the QM/MM dimer configuration, which means that the isotropic damping prevents the accumulation of charge between the two \ce{H20} molecules. Moving on to the 12th image, the excess charge between the QM and the MM \ce{H2O} increases. Eventually, the non-damped calculation undergoes a significant decrease in the potential energy, illustrating the polarization catastrophe.
Hence, implementing an isotropic damping function can prevent the decline of potential energy, averting the polarization catastrophe.

\subsection{Graphene-Water Interface}
The local water structure of the pure QM and PE-QM/MM MD simulation above the graphene sheet is analyzed using $z$ distribution functions. These distributions are obtained by calculating the distances between the O atoms of the \ce{H2O} molecules and the graphene sheet. For the QM and QM/MM simulation, the $g_{\mathrm{G,O}(z)}$ exhibits distinct bands for the first and second water solvation layers. The bands of the third and fourth water solvation layer are broadened. Thus, the \ce{H2O} molecules closer to the surface experience stronger coordination, enabling a more rigid water structure compared to the third and fourth water solvation layers.
Notably, the first band shows a significant shoulder in the QM MD simulation, which is visible in the QM/MM simulation as well. 
Above 6.0~\AA\, in the MM region, the $g_{\mathrm{G,O}(z)}$ of the QM/MM simulation matches the bands of the full QM simulation. In particular, the second band shows a good agreement. The bands further away become noisier and slightly shifted compared to the full QM distribution. 
The $g_{\mathrm{G,O}(z)}$ of the QM and QM/MM simulations illustrate similar positions of the bands and dips, indicating a good agreement between both simulations. The difference between the $g_{\mathrm{G,O}(z)}$ is mainly due to insufficient sampling of the MD simulation. 

For the MD simulations, a time step of 1.0 fs in our SAFIRES Langevin molecular dynamics integrator, with the temperature kept constant at 300 K. The SAFIRES boundary is anchored to the \ecom\ of the outermost QM H$_2$O molecule. The internal geometry of both the QM and MM H$_2$O molecules 
are constrained to the ground state geometry of a 
PBE H$_2$O molecule. 
The RDF is sampled every 50 fs, from an overall 100 ps simulation window.
\begin{figure}[t]
\centering
\includegraphics[width=\linewidth]{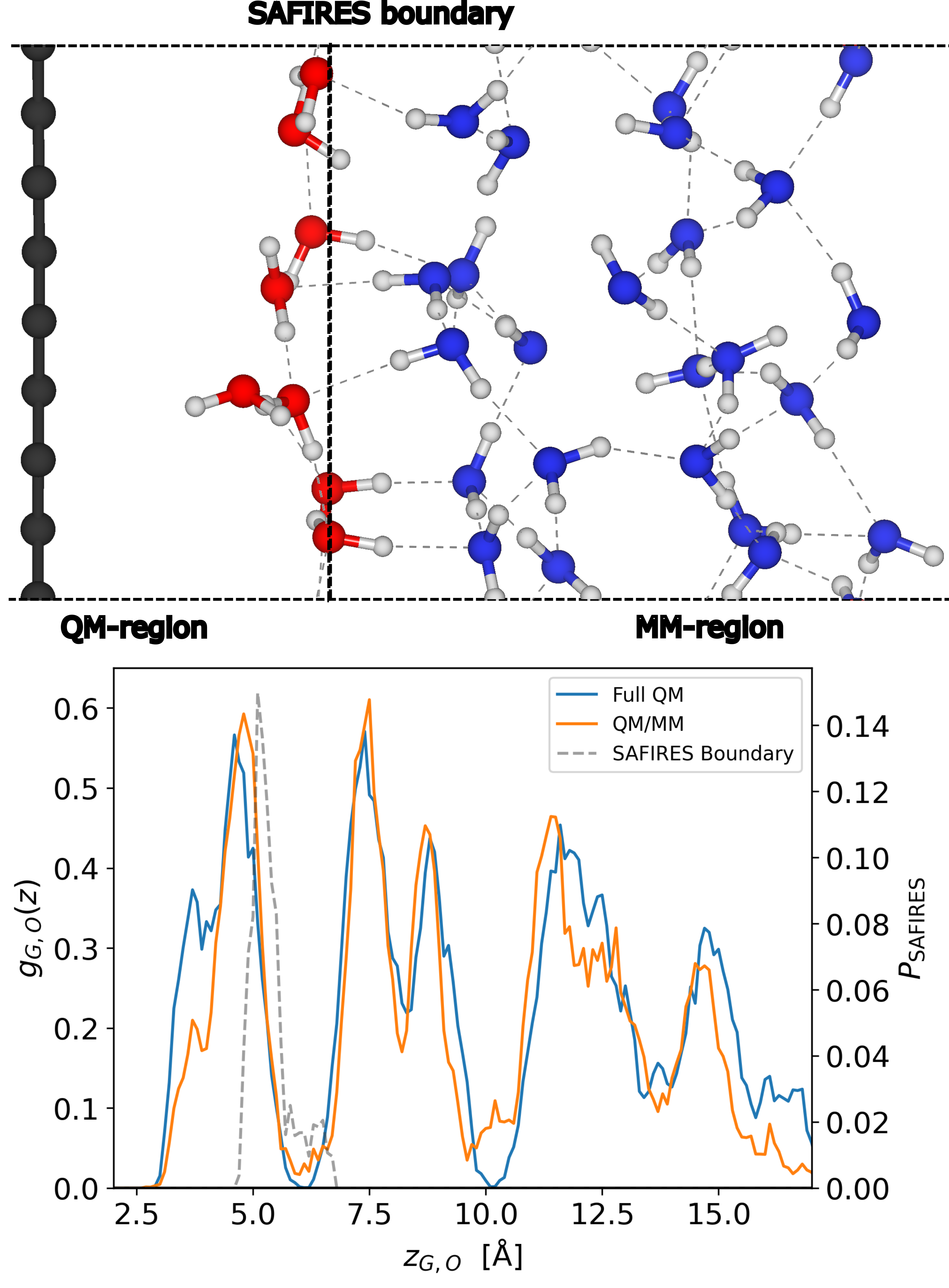}
\caption{Molecular dynamics simulation of a solvated graphene with four layers of \ce{H2O} molecules (top). The first \ce{H2O} layer is described with the PBE energy functional, and the rest of the layers with the SCME potential function - parameterized with PBE moment tensors. The left and right hand side of the QM subsystem is padded with 5.5 \AA\ of vacuum. The PBE water layer is composed of 8 \ce{H2O} molecules, leaving the MM region with 24 \ce{H2O} molecules.
The graphene-oxygen distribution function, $g_{G,O}(z)$, as a function of the non-periodic $z$-axis is plotted (bottom) for a MD simulation of the pure QM system (blue), and the QM/MM system described above (orange). The position and relative propability of the SAFIRES boundary is plotted (grey dashed lines), showing that the boundary is localized at the top of the QM \ce{H2O} layer -- with some attempted crossing between layers, as indicated the non-zero probability in the region between the layers.}
\label{fig:MD_Graphene}
\end{figure}

\section{Conclusion \& Outlook}
We illustrated a polarizable embedding QM/MM scheme with an isotropic damping function to provide reliable QM/MM calculations for 2D periodic systems. Based on the Kohn-Sham DFT and the SCME model of \ce{H2O} molecules, we enable a mutual polarization calculation scheme, handling large-scale effects at the solid/liquid interface. 
First, we investigated 2D periodic systems using ice layers with a layered and a mixed QM/MM configuration. In both cases, we showed that the QM/MM interaction energy converges to the interaction energy of the pure QM calculation. Beyond that, the 2D periodicity is maintained via the inner and outer cell grid expansion. The latter introduces an efficient and accurate way of sampling long-range electrostatics, using the origin shift to handle multipole moments in cells far away. With the help of the outer grid expansion $Nc_{\mathrm{outer}}$, a fast convergence with a low inner grid expansion of $Nc$ = [1,1,0] and a high outer cell grid expansion of $Nc_{\mathrm{outer}}$ = [9,9,0] can be enabled. Noteworthy, the convergence pattern is independent of the real-space grid and the QM/MM configuration. Besides that, employing an outer cell grid expansion can boost the PE-QM/MM simulation by a factor of five.
The origin shift provides a beneficial possibility to calculate the multipole moments in the outer cells.
Thus, the PE-QM/MM simulation can describe the energetics, but also the potential in the solid part of the QM region with high accuracy comparable to a pure QM calculation.
Further, we implemented an isotropic damping function for sites nearby the QM/MM boundary to handle the polarization catastrophe. Hence, the electrostatic interaction between QM and MM sites at the boundary is damped. Using the gold-water PE-QM/MM MD simulation, we tested the isotropic damping function, which exhibits a significant effect on the boundary molecules. 
Moreover, the PE-QM/MM simulation was employed on a graphene-water system to examine the water distribution along the QM and MM regions. According to the RDF, the PE-QM/MM MD showed similar results compared to the pure QM MD simulation, indicating that our PE-QM/MM approach enables a reasonable description of the solid/liquid interface.
As a next step, the PE-QM/MM scheme can be coupled with a grand-canonical DFT approach to mimic the electrode potential. Thereby, the slab in the QM region can polarize the solvent in the MM region and vice versa. As a result, the PE-QM/MM approach provides a general and realistic simulation for electrochemical systems.

\begin{acknowledgments}
We thank Prof. Marko Melander and Dr. Yorick Schmerwitz for fruitful discussions. 
This work was supported by the Icelandic Research Fund, grant no.\ 2410644. Computer resources, data storage, and user support were provided by the 
Icelandic Research e-Infrastructure (IREI), funded by the Icelandic Infrastructure Fund.
\end{acknowledgments}

\bibliography{main}

\appendix

\section{Isotropic Real-Space Damping Functions}
\label{app:isotropic}

The isotropic real-space damping relies on spatially resolved molecular centered functions - which are normalized such that $\int \sum_a w^a(\br^a,\br)d\br = 1$.
For this purpose, we choose normalized Gaussians and define molecular densities:
\begin{equation}
    n^{a}(\bfr^a,\bfr) =\begin{cases} 
\frac{1}{\gamma \sqrt{2\pi}} e^{-|\bfr - \bfr^a|^2/2\gamma^2}\ \ \mathrm{for}\ |\bfr-\bfr^a| < R_{c,\mathrm{n}}, \\
0 \ \mathrm{for}\ |\bfr-\bfr^a| \geq R_{c,\mathrm{n}}
    \end{cases}
    \label{eq:molDens}
\end{equation}
where the cut-off $R_{c,\mathrm{n}}$ is chosen such that $S^G$, Eq.~\eqref{eq:S_G_beta}, goes smoothly to zero.
With this we construct a weight function around each QM center as
\begin{equation}
    w^a(\bfr^a,\bfr) = \frac{n^a(\bfr) + n^\mathrm{rest}(\bfr)}{n^\mathrm{tot}(\bfr)}
\end{equation}
where the total molecular density is given by
\begin{equation}
    n^\mathrm{tot}(\bfr^a, \bfr) = \sum_bn^b(\bfr^b,\bfr) + n^\mathrm{rest}(\bfr)
\end{equation}
The rest density is
\begin{equation}
    n^\mathrm{rest}(\bfr) =\begin{cases} 1\forall \bfr \in \sum_bn^b(\bfr)=0 \\
0\end{cases}
\end{equation}
and is simply a space filler added due to the cut-off applied to the molecular densities in Eq.~\eqref{eq:molDens}. 
It can be written as
\begin{equation}
    n^\mathrm{rest} = \prod_a \delta(|\bfr - \bfr^a|-R_{c,\mathrm{n}})
\end{equation}
This effective spatially resolved damping scheme is shown schematically in FIG.~\ref{fig:isodamp}.
The distance between the two centers of mass controls the magnitude of the damping, $S^\mathrm{G}$, which is applied in a spherical region surrounding the QM water molecule - whose radius is controlled by $R_{c,\mathrm{n}}$.

In the case of an arbitrary number of QM centers we modify the molecular densities and directly incorporate the cut-off $R_{c,\mathrm{n}}$ such that
\begin{equation}
n^{a'}(\bfr) = n^a(\bfr)\Theta(R_{c,\mathrm{n}} - |\bfr - \bfr^a|)
\end{equation}
resulting in
\begin{align}
w^a(\bfr^a,\bfr) =& \frac{n^{a'} + n^\mathrm{rest}(\bfr)}{n^{a'} + \sum_{b\neq a}n^{b'} + n^\mathrm{rest}(\bfr)} \nonumber \\
=& \frac{f(x)g(x) + h(x)}{f(x)g(x) + h(x) + W}
\end{align}
where $W = \sum_{b\neq a}n^{b'}$.
The derivative becomes
\begin{align}
\frac{\partial w^a(x)}{\partial x} =& W\frac{f'(x)g(x) + f(x)g'(x)}{(f(x)g(x) + h(x) + W)^2} \nonumber \\ +& W\frac{h'(x)}{(f(x)g(x) + h(x) + W)^2}
\end{align}

\section{Forces at Self-Consistency}
The forces due to the isotropic damping function acting on the MM and QM nuclei are derived in the case of a dipole plus induced dipole only. Generalization to higher order moments and polarizabilities is relatively straightforward, and the rest of the force expressions can be found in the Supporting Information S1.

In this case the energy functional for the explicit coupling is
\begin{align}
E^{\rQMMM}[\rho,M'] =& \int \rho(\mathbf{r})V^\mathrm{MM}(\mathbf{r})d\mathbf{r} \nonumber \\
=& \sum_i \int  S^i(\bfr)\rho(\bfr)T^{ri,d}_\alpha(\mu^i_\Ga + \Delta\mu^{i}_{\Ga})d\bfr \nonumber \\
=& \sum_i(\mu^i_\Ga + \Delta\mu^{i}_{\Ga})V^{i\rQM}_\Ga[S^{i}]
\end{align}
with self-energy 
\begin{align}
E^{\rself}[\rho(\bfr)] =& -\frac{1}{2}\sum_{i}\Dmu^{i}_\Ga[S^{i}] V^i_\Ga[S^{i}] \nonumber \\ =& -\frac{1}{2}\sum_i\alpha^i_{\Gab}V^i_{\Ga}[S^{i}]V^{i}_{\Gb}[S^{i}]
\end{align}
At SCF we satisfy
\begin{equation}
 \frac{\partial E_\rself}{\partial \Delta \mu^i_\al} =0,\ \ \frac{\partial V^i_\beta}{\partial \Delta \mu^i_\al}=0
\end{equation}
therefore
\begin{equation}
 F^{i}_\Ga = -\frac{\partial E^{\rQMMM}}{\partial S^{i}}\frac{\partial S^{i}}{\partial r^i_\Ga}
\end{equation}
More explicitly these terms are
\begin{equation}
F^i_\Ga = \int \left(\frac{\partial S^i(\bfr)}{\partial \bfr^i_\Ga}V^i(\bfr) + S^i(\bfr)\frac{\partial V^{i'}(\bfr)}{\partial \bfr^i_\Ga}\right)\rho(\bfr)d\bfr
\end{equation}
where
\begin{equation}
\frac{\partial S^i(\bfr)}{\partial r^i_{\Ga}} = S'(|\bfr^a-\bfr^i|)w^a(\bfr,\bfr^a)
\end{equation}
The force per MM atom becomes
\begin{equation}
F^{i,k}_\Ga = \frac{m_k}{\sum_{j\in i} m_j}F^{i}_\Ga
\end{equation}
and for the QM center of mass we have
\begin{equation}
F^a_\Ga = \int \frac{\partial S^i(\bfr)}{\partial \bfr^a_\Ga}V^i(\bfr)\rho(\bfr)d\bfr
\end{equation}
where
\begin{align}
\frac{\partial S^i(\bfr)}{\partial r^a_{\Ga}} =& S'(|\bfr^a-\bfr^i|)w^a(\bfr,\bfr^a) \nonumber \\ 
+& S(|\bfr^i - \bfr^a|)w^{'a}(\bfr,\bfr^a)
\end{align}
resulting in force per atom
\begin{equation}
F^{a,k}_\Ga = \frac{m_k}{\sum_{j \in a} m_j}F^{a}_\Ga
\end{equation}
In the limit of a single QM center it is easy to show that $\nabla w^{a}(\bfr^a,\bfr) = 0$. In fact all partial derivatives of the weight function with respect to the position of the QM center vanish at the boundary between $n^a(\bfr)$ and $n^\mathrm{Rest}(\bfr)$.\\

\end{document}


The details of the PAW Hamiltonain, derivation of Forces, self-energies and interaction tensor damping functions can be in the Supplementary Information in Jonsson et. al. \cite{jonsson2019polarizable}.

\section{PAW Hamiltonian and Forces}

Before deriving the Hamiltonian and forces acting on the QM nuclei, the projector augmented wave (PAW) theory\cite{paw1} is first reviewed briefly.
In PAW, the rapidly oscillating 
KS--DFT wave functions (wfs), $\psi_m(\br)$, are transformed to 
smooth pseudo wave functions (pseudo-wfs), $\tpsi_m(\br)$, near the atomic core region. 
Within this framework the pseudo-wfs can be represented on a (relatively) coarse 
real-space grid, $\tpsi_m(\br)\rightarrow\tpsi_m(\br^G)$, and the corresponding 
pseudo electron density, $\tin(\br)$, is represented on a fine grid,
$\tin(\br)\rightarrow\tin(\br^g)$
(typically eight times the size of the coarse grid). Calculations can be
performed quite efficiently since the mean-field effective Hamiltonian 
is first evaluated accurately in terms of the fine grid pseudo electron 
density, and in a second step transformed to the coarse grid representation, 
$H^\mr{KS}_\mr{eff}(\br^g)\rightarrow H^\mr{KS}_\mr{eff}(\br^G)$. 
The coarse grid effective Hamiltonian is used
when solving the KS--DFT eigenvalue problem. 
The interested reader is
referred to the original PAW work by Blochl\cite{paw1},
and the implementation in GPAW\cite{GPAW1}, for more details.

The transformation from the wfs to the pseudo-wfs is achieved 
with a transformation operator
\begin{equation}
 \mathcal{T} = 1 + \sum_a\sum_n\left(\ket{\phi_n^a}
                 -\ket{\ti{\phi}_n^a}\right)\bra{\ti{p}^a_n}
\end{equation}
where $\ti{\phi}^a_n$ are smooth atomic centered partial waves, and
$\phi^a_n$ are the 
original, rapidly oscillating atom centered 
partial waves. Beyond a cut-off radius $r^a_c$ they are identical
\begin{equation}
 \phi^a_n(\br) = \ti{\phi}^a_n(\br), \qquad |\br - \br^a| > r^a_c
\end{equation}
and in turn this cut-off radius defines an atomic centered augmentation
sphere around nucleus $a$. The atomic centered projectors, $\ti{p}^a_n$
are chosen such as to form a complete dual basis to the pseudo partial waves
\begin{equation}
 \sum_n\ket{\ti{p}^a_n}\bra{\ti{\phi}^a_n} = 1, 
 \qquad \braket{\ti{p}^a_n|\ti{\phi}^a_l} = \delta_{nl}
\end{equation}
The wfs can be defined in terms of the pseudo-wfs as
\begin{align}
 \ket{\psi_m} = \ket{\tpsi_m} + \sum_a\sum_n\left(\ket{\phi_n^a}
                 -\ket{\ti{\phi}_n^a}\right)\braket{\ti{p}^a_n|\tpsi_m}
\end{align}
and practical calculations carried out with the pseudo-wfs. In terms of the 
pseudo-wfs and projectors, atomic density matrices can be defined as
\begin{equation}
 D^a_{nl} = \sum_mf_m\braket{\ti{p}^a_n|\tpsi_m}\braket{\tpsi_m|\ti{p}^a_l}
\end{equation}
where $f_m$ is the occupation number of KS--DFT state $m$, 
and the pseudo electron density defined as
\begin{equation}
 \tin(\br) = \sum_mf_m\braket{\tpsi_m|\tpsi_m} + \sum_a\tin^a_c(\br)
\end{equation}
where $\tin^a_c$ is a smooth pseudo core density, equal to the
core density outside of the augmentation sphere. The all-electron 
density is defined as
\begin{equation}
 n(\br) = \tin(\br) + \sum_a\left[n^a(\br) - \tin^a(\br)\right]
\end{equation}
where the atomic corrections are given by
\begin{align} 
 \left[n^a(\br) - \tin^a(\br)\right] =& 
 \sum_{nl}D^a_{nl}\left(\phi^{a*}_n(\br)\phi^a_l(\br) -
 \ti{\phi}^{a*}_n(\br)\ti{\phi}^{a}_l(\br)\right) \nonumber \\
 &+ \left(n^a_c(\br) - \tin^a_c(\br)\right) \label{eq:paw1}
\end{align}
with $n^a_c(\br) = \sum_i\braket{\phi_{c,i}^a|\phi_{c,i}^a}$ -- where $\phi_{c,i}^a$
are describing the core electron states. Complementary to the pseudo-electron density
is a pseudo-charge density
\begin{equation}
 \ti{\rho}(\br) = \tin(\br) + \sum_a\ti{Z}^a(\br) \label{eq:pseudocd}
\end{equation}
where $\ti{Z}^a(\br)$ are atomic compensation charges. The
compensation charges are defined in terms of multipole moments $Q^a_L$
\begin{equation}\label{eq:paw2}
 \ti{Z}^a(\br) = \sum_LQ^a_L\ti{g}^a_L(\br)
\end{equation}
where $\ti{g}^a_L(\br) = \ti{g}^a_l(\br)Y_L(\hat{\br})$ is a shape function
written in terms of spherical harmonics $Y_L(\hat{\br})$ and is bound 
to fulfill $\int r_l\ti{g}^a_l(\br)Y_L(\hat{\br})d\br = 1$. L is a combined
angular ($l$) and magnetic quantum number. An additional constraint 
is applied by requiring that the electron density and pseudo-electron 
density have identical multipole moments outside the augmentation sphere.
This can be expressed as 
\begin{equation}\label{eq:paw3}
 \int r_l\left[\tin(\br) + \ti{Z}^a(\br) 
      - n^a(\br) - Z^a(\br)\right]Y_L(\hat{\br})d\br = 0
\end{equation}
where $Z^a(\br) = \mathcal{Z}^a\delta(\br - \br^a)$ is the atomic number of nuclei $a$.
Inserting equations \eqref{eq:paw1} and \eqref{eq:paw2} into equation \eqref{eq:paw3}
and solving for the multipole moments results in
\begin{equation}
 Q^a_L = \Delta^a\delta_{l0} + \sum_{nl}\Delta^a_{L,nl}D^a_{nl}
\end{equation}
The constants $\Delta^a\delta_{l0}$ and $\Delta^a_{nl}$ depend on the
atomic number of the nuclei and the pre-defined partial waves
and frozen core electron density, as well as the pseudo-partial waves
and pseudo-core electron density, all of which is calculated and stored 
beforehand. 
The main reason for this definition of the atomic
compensation charges is that it makes the evaluation of the 
KS--DFT Coulomb term relatively straightforward.

For the purpose of electrostatic interactions in a QM/MM interface 
we can also make use the pseudo-charge density
to derive both the appropriate PAW QM/MM Hamiltonian
and the atomic forces.\cite{dohn2017grid} 
As long as the MM sites do not overlap with 
the PAW augmentation spheres the electrostatic interactions outside of 
the sphere are accurately described by virtue of equation
\eqref{eq:paw3}. This means that in practical applications of PAW based QM/MM
interfaces the electrostatic interaction between the QM and MM subsystems is
\begin{equation}
 \int\ti{\rho}(\br)V^\mr{QM/MM}_\mr{ext}(\br)d\br = E^\mr{QM/MM}_\mr{ext}
\end{equation}
The PAW QM/MM Hamiltonian can be obtained from the relation
\begin{equation}
 \frac{\partial E^\mr{QM/MM}_\mr{ext}}{\partial \bra{\tpsi_m}} = 
 f_m\ti{H}^\mr{QM/MM}_\mr{ext}\ket{\tpsi_m}
\end{equation}
which gives
\begin{align}
 \frac{\partial E^\mr{QM/MM}_\mr{ext}[\ti{\rho}(\br)]}{\partial \tpsi^*_m(\br')} =&
 \frac{\partial E^\mr{QM/MM}_\mr{ext}[\tin(\br) + \sum_a\ti{Z}^a(\br)]}{\partial \tin(\br)}
 \frac{\partial \tin(\br)}{\partial \tpsi^*_m(\br')} \nonumber \\
 &+\sum_a\sum_{nl}\int d\br 
 \frac{\partial E^\mr{QM/MM}_\mr{ext}[\tin(\br) + \sum_a\ti{Z}^a(\br)]}{\partial \ti{Z}^a(\br)}
 \frac{\partial \ti{Z}^a(\br)}{\partial D^a_{nl}}\frac{\partial D^a_{nl}}{\partial \tpsi^*_m(\br')}
 \nonumber \\
 =& \int V^\mr{QM/MM}_\mr{ext}(\br)\delta(\br - \br')\tpsi_m(\br)d\br \nonumber \\
 &+ \sum_a\sum_{nl}
 \left[\int V^\mr{QM/MM}_\mr{ext}(\br)\sum_L \Delta^a_{L,nl}\ti{g}^a_L(\br)d\br \right]
 f_m\ti{p}^a_{n}(\br')\braket{\ti{p}^a_l|\tpsi_m}
\end{align}
resulting in 
\begin{equation}
 \ti{H}^\mr{QM/MM}_\mr{ext} = V^\mr{QM/MM}_\mr{ext}(\br) 
 + \sum_a\sum_{nl}\ket{\ti{p}^a_n}\Delta V^{\mr{QM/MM},a}_{\mr{ext},nl}\bra{\ti{p}^a_l} .
\end{equation}
The force on atom $z$ in the QM region due to the QM/MM electrostatic plus induction
energy is 
\begin{align}
 F^{a,\mr{QM/MM}}_\al =& -\frac{\partial}{\partial r^a_\al}
 \int\ti{\rho}(\br)V^\mr{QM/MM}_\mr{ext}(\br)d\br \nonumber \\
 =& -\int V^\mr{QM/MM}_\mr{ext}(\br)\frac{\partial \ti{\rho}(\br)}{\partial \tin(\br)}
 \frac{\partial \tin(\br)}{\partial r^a_\alpha} d\br 
 -\int\sum_L V^\mr{QM/MM}_\mr{ext}(\br)
 \frac{\partial \ti{\rho}(\br)}{\partial \ti{g}_L^a(\br)}
 \frac{\partial \ti{g}_l^a(\br)}{\partial r^a_\alpha}d\br \nonumber \\
 &-\sum_{nl}\int V^\mr{QM/MM}_\mr{ext}
 \frac{\partial \ti{\rho}(\br)}{\partial D^a_{nl}}
 \frac{\partial D^a_{nl}}{\partial r^a_\alpha}d\br \nonumber \\
 =& -\int V^\mr{QM/MM}_\mr{ext}(\br)\frac{\partial \ti{n}^a_c(\br)}{\partial r^a}d\br
 -\int\sum_L V^\mr{QM/MM}_\mr{ext}(\br)Q^a_L
 \frac{\partial \ti{g}^a_L(\br)}{\partial r^a_\alpha} \nonumber \\
 &- \sum_{nl}\Delta V^{\mr{QM/MM},a}_{\mr{ext},nl}\frac{\partial D^a_{nl}}{\partial r^a_\alpha} .
\end{align}


\section{Tensor Damping Functions}

The position of MM molecules in the global reference 
frame will possibly place them within the QM grid space, and hence 
close to or on top of a grid point. Potentials and gradients in 
terms of the interaction tensors will diverge resulting in 
what is commonly known as the polarization catastrophe 
-- as coined by Thole\cite{thole:1981}.
In order to avoid this, tensor damping functions are included in 
the formulation. Effectively, the functions smear out
the point moments and describe a screened interaction as the 
electron densities
start to overlap.

Common choices include Thole type\cite{thole:1981} damping 
functions\cite{masia2005,masia2006,burnham1999}
-- based on an exponential decay description of point charges -- 
or a normalized Gaussian description\cite{Stone:2011}. A more 
comprehensive comparison damping functions is beyond the 
scope of this work but can be found 
elsewhere.\cite{dampingbothyeah,masia2005,masia2006} Gaussian type 
damping functions are used in this work.

In order to arrive at the appropriate screening -- or damping -- 
function for the various orders of the interaction tensors the zeroth
order tensor is modified as follows 
\begin{equation}
 T^{ij,d} = \frac{1}{r}\lambda_0(r)
\end{equation}
where
\begin{equation}\label{eq:gaussdamp}
 \lambda_0(r) = \mr{erf}\left(\frac{r}{g}\right)
\end{equation}
for normalized Gaussian type damping. The damping function subscript 
denotes the order of the tensor it applies to. Evaluating the 
gradient results in
\begin{align}
 T^{ij,d}_\al =& -\frac{r_\al}{r^3}\lambda_1(r) \\
 T^{ij,d}_\ab =& 3\frac{r_\al r_\be}{r^5}\lambda_2(r) -
 \frac{\delta_\ab}{r^3}\lambda_1(r) \\
 \vdots&  \nonumber 
\end{align}
It is easy to show that starting from equation \eqref{eq:gaussdamp} 
the next order damping function -- in the series of orders $\{1,2,3,4\dots\}$ 
-- can be evaluated as
\begin{equation}
 \lambda_{k}(r) = \lambda_{k-1}(r) - 
 \frac{r}{2k-1}\frac{\partial}{\partial r}\lambda_{k-1}(r)
\end{equation}


For clarity $\left(\frac{r}{g}\right)$ is denoted as $S$ -- screened 
distance. The Gaussian type damping functions are:
 \begin{align}
 \lambda_1(r) =&\ \mr{erf}(S) - \frac{2}{\sqrt\pi}Se^{-S^2} \\
 \lambda_2(r) =&\ \mr{erf}(S) - \frac{2}{\sqrt\pi}\left(S + \frac{2}
 {3}S^3\right)e^{-S^2} \\
 \lambda_3(r) =&\ \mr{erf}(S) - \frac{2}{\sqrt\pi}\left(S + \frac{2}
 {3}S^3 + \frac{4}{15}S^5\right)e^{-S^2} \\
 \lambda_4(r) =&\ \mr{erf}(S) - \frac{2}{\sqrt\pi}\left(S + \frac{2}
 {3}S^3 + \frac{4}{15}S^5
+ \frac{8}{105}S^7\right)e^{-S^2} \\
 \lambda_{5}(r) =&\ \mr{erf}(S) - \frac{2}{\sqrt\pi}\left(S + 
 \frac{2}{3}S^3 + \frac{4}{15}S^5
                           + \frac{8}{105}S^7 + \frac{16}{945}S^9\right)e^{-S^2} 
\end{align}
Another formula to evaluate a particular order of the Gaussian type damping function is 
\begin{equation}
    \lambda_\mr{n}(r) = \mr{erf}(S) - \frac{2}{\sqrt{\pi}}
    \left(\sum_{i=1}^n S^{2(i-1) + 1}\frac{2^{i-1}}{\prod_{j=1}^i (2j-1)} \right)e^{-S^2}
\end{equation}
where $n$ is the order of the interaction tensor.

The complete set of damped interaction tensors describing all electrostatic 
interactions in the QM/MM interface are then:
\begin{align}
  \Ta^{ij,d} =&\ -\frac{r_\al}{r^3}\lambda_1(r) \\
 \Tab^{ij,d} =&\ 3\frac{r_\al r_\be}{r^5}\lambda_2(r) 
 - \frac{\delta_\ab}{r^3}\lambda_1(r) \\
 \Taby^{ij,d} =&\ -15\frac{r_\al r_\be r_\gamma}{r^7}\lambda_3(r) 
            + \frac{3}{r^5}\left(r_\al\delta_\by 
            + r_\be\delta_{\alpha\gamma} 
            + r_\gamma\delta_\ab\right)\lambda_2(r) \\ 
 \Tabyd^{ij,d} =& 105\frac{r_\al r_\be r_\gamma r_\delta}{r^9}\lambda_4(r) 
 - \frac{15}{r^7}\begin{pmatrix}
   r_\al r_\be\delta_{\gamma\delta} 
   + &\!\!\!\! r_\al r_\gamma\delta_{\be\delta} 
   + &\!\!\!\! r_\al r_\delta\delta_{\be\gamma} \\
   + r_\be r_\gamma\delta_{\al\delta} 
   + &\!\!\!\! r_\be r_\delta\delta_{\al\gamma} 
   + &\!\!\!\! r_\gamma r_\delta\delta_{\ab}                             
      \end{pmatrix}\lambda_3(r) \nonumber \\
      &+\frac{3}{r^5}\left(\delta_\ab\delta_{\gamma\delta}
       +\delta_{\al\gamma}\delta_{\be\delta}
       +\delta_{\al\delta}\delta_{\be\gamma}\right)\lambda_2(r)
\end{align}

\begin{align}
     \Tabyde^{ij,d} =& 
 -945\frac{ r_\al r_\be r_\gamma r_\delta r_\epsilon}
          {r^{11}}\lambda_{5}(r) \nonumber \\
  + \frac{105}{r^9}&\begin{pmatrix}
    r_\al r_\be r_\gamma\delta_{\delta\epsilon} 
   + &\!\!\!\! r_\al r_\be r_\delta\delta_{\gamma\epsilon} 
   + &\!\!\!\!
    r_\al r_\be r_\epsilon\delta_{\gamma\delta} 
   + &\!\!\!\! r_\al r_\gamma r_\delta\delta_{\be\epsilon} 
   + &\!\!\!\! 
    r_\al r_\gamma r_\epsilon\delta_{\be\delta} \\
   + r_\al r_\delta r_\epsilon\delta_{\be\gamma} 
   + &\!\!\!\! r_\be r_\gamma r_\delta\delta_{\al\epsilon} 
   + &\!\!\!\!
    r_\be r_\gamma r_\epsilon\delta_{\al\delta} 
   + &\!\!\!\! r_\be r_\delta r_\epsilon\delta_{\al\gamma} 
   + &\!\!\!\!
    r_\gamma r_\delta r_\epsilon\delta_{\al\be} 
                    \end{pmatrix}\lambda_4(r) \nonumber \\
   -\frac{15}{r^7}&\begin{pmatrix}
  r_\al\delta_{\be\gamma}\delta_{\delta\epsilon} 
 + &\!\!\!\! r_\al\delta_{\be\gamma}\delta_{\gamma\epsilon} 
 + &\!\!\!\!
  r_\al\delta_{\be\epsilon}\delta_{\gamma\delta} 
 + &\!\!\!\! r_\be\delta_{\al\gamma}\delta_{\delta\epsilon} 
 + &\!\!\!\!
  r_\be\delta_{\al\delta}\delta_{\gamma\epsilon} \\
 + r_\be\delta_{\al\epsilon}\delta_{\gamma\delta} 
 + &\!\!\!\! r_\gamma\delta_{\al\be}\delta_{\gamma\epsilon} 
 + &\!\!\!\!
  r_\gamma\delta_{\al\delta}\delta_{\be\epsilon} 
 + &\!\!\!\! r_\gamma\delta_{\al\epsilon}\delta_{\be\delta} 
 + &\!\!\!\!
  r_\delta\delta_{\al\be}\delta_{\gamma\epsilon} \\ 
 + r_\delta\delta_{\al\gamma}\delta_{\be\epsilon} 
 + &\!\!\!\! r_\delta\delta_{\al\epsilon}\delta_{\be\gamma} 
 + &\!\!\!\!
  r_\epsilon\delta_{\al\be}\delta_{\gamma\delta} 
 + &\!\!\!\! r_\epsilon\delta_{\al\gamma}\delta_{\be\delta} 
 + &\!\!\!\!
  r_\epsilon\delta_{\al\delta}\delta_{\be\gamma}
                   \end{pmatrix}\lambda_3(r)  .
\end{align}

Potentials, fields and field gradients are then modified accordingly (and by extension energy and forces), for example 
\begin{align}
  V^{i\mr{QM}}_\al =& \int\rho_\mr{QM}(\br)\left(-\frac{r_\al}{r^3}\lambda_1(r)\right)d\br
   = \int\rho_\mr{QM}(\br)\Ta^{ir,d} d\br \\
  V^{i\mr{QM}}_\ab 
  =& \int \rho_\mr{QM}(\br)\left(3\frac{r_\al r_\be}{r^5}\lambda_2(r)
  - \frac{\delta_\ab}{r^3}\lambda_1(r)\right)d\br = \int\rho_\mr{QM}(\br)\Tab^{ir,d} d\br \label{eq:secondorderpot}
\end{align}


\section{Self-Energy and Forces}
At self-consistency of the PE-QM/MM interface (and also at self-consistency for the MM part only - Single Center Multipole Expansion model section in the main text), the MM induced dipoles and quadrupoles can be written as
\begin{align}
\Delta\mi_\al = -\alpha_\ab V^i_\be 
                - \frac{1}{3}A_{\al,\by}V^i_\by = \Delta\mi_\al(\al) + \Delta\mi_\al(A)\\
\Delta\qi_\ab = -A_{\gamma,\ab}V^i_\gamma 
                   - C_{\gamma\delta,\ab}V^i_{\gamma\delta} = \Delta\qi_\ab(A) + \Delta\qi_\ab(C)
\end{align}
where on the right hand side of the second equality the contribution from the external field and field gradient due to the on-site potential is split up. 
With these definitions it is easy to relate the external field and field gradient at site $i$ to the self-consistent moments of 
molecule $i$
\begin{align}\label{eq:indtopot}
    V^i_\be =& -\frac{\Delta\mi_\al(\al)}{\al_\ab} \\
    V^i_\gamma =& -\frac{\Delta\qi_\ab(A)}{A_{\gamma,\ab}} \label{eq:indtopotquad}
\end{align}
and
\begin{align}\label{eq:indtopotgrad}
    V^i_{\gamma\delta} =& -\frac{\Delta\qi_\ab(C)}{C_{\gamma\delta,\ab}} \\
    V^i_\by =& -\frac{\Delta\mi_\al(A)}{A_{\al,\by}} \label{eq:indtopotgraddip} .
\end{align}
The self-energy in linear response theory is 
\begin{equation}
    E_\mr{self}(1) = -\int_{0}^{\Delta\mi} V^i_\be d\Delta\mi .
\end{equation}
It gives the energy cost of inducing a first order moment in the potential field at site $i$. By inserting the relation in equation \eqref{eq:indtopot} into the equation above, and by considering only (for the moment) the induced dipole in in response to an external field gives
\begin{equation}
    E_\mr{self}(1) = \int_0^{\Delta\mi(\al)}\frac{\Delta\mi_\al(\al)}{\al_\ab}d\Delta\mi 
    = \frac{1}{2}\frac{\Delta\mi_\al(\al)\Delta\mi_\be(\al)}{\al_\ab} .
\end{equation}
For isotropic atomic polarization this becomes
\begin{equation}
    E_\mr{self}^\mr{iso} = \frac{1}{2}\frac{(\Delta\mi)^2}{\al} .
\end{equation}
This form is most frequently encountered in PE-QM/MM work based on isotropic atomic polarization and induced dipole in response to the external field.
Similarly, for the induced quadrupole 
\begin{equation}
    E_\mr{self}(2) = -\frac{1}{3}\int_0^{\Delta\qi} V^i_\by d\Delta\qi
\end{equation}
which expresses the energy cost of inducing a second-order moment in the field gradient at site $i$. The factor of $1/3$ follows from the definition of the traceless Cartesian moments\cite{stone2013theory} used in SCME. The total self-energy for a single site $i$ is then
\begin{align}
    E^\mr{tot}_\mr{self} =& E_\mr{self}(1) + E_\mr{self}(2) \nonumber \\
    =& \int_0^{\Delta\mi}\frac{\Delta\mi_\al(\al)}{\al_\ab}d\Delta\mi 
    + \frac{1}{3}\int_0^{\Delta\qi}\frac{\Delta\qi_\ab(C)}{C_{\gamma\delta,\ab}}d\Delta\qi \nonumber \\
    =& \frac{1}{2}\frac{\Delta\mi_\al(\al)}{\al_\ab}\left(\Delta\mi_\be(\al) 
    +\frac{1}{3}\Delta\mi_\be(A)\right) + \frac{1}{6}\frac{\Delta\qi_\ab(C)}{C_{\gamma\delta,\ab}}\left(\Delta\qi_{\gamma\delta}(C)
    +\Delta\qi_{\gamma\delta}(A)\right) \nonumber \\
    =& \frac{1}{2}\frac{\Delta\mi_\al(\al)\Delta\mi_\be(\al)}{\al_\ab} 
     + \frac{1}{3}\frac{\Delta\mi_\al(\al)\Delta\qi_\by(C)}{k_{\al,\by}}
     + \frac{1}{6}\frac{\Delta\qi_\ab(C)\Delta\qi_{\gamma\delta}(C)}{C_{\gamma\delta,\ab}}
     \label{eq:selfenergy}
\end{align}
where the relations in equations \eqref{eq:indtopot}--\eqref{eq:indtopotgraddip} are used. The matrix $k$ is given by
\begin{equation}
 k = \frac{\al C}{A}
\end{equation}
This expression for the self-energies is very useful at self-consistency of the PE-QM/MM interface, or for pure MM calculations. 

Considering, as an example, the total electrostatic plus induction energy functional in equation (11) in the main text
\begin{align}
 E^\mr{MM}_\mr{ele+ind} =& \frac{1}{2}\sum^{n_\mr{MM}}_i
 \left((\mi_\al+\Delta\mi_\al)V^i_\al
 + \frac{1}{3}(\qi_\ab + \Delta\qi_\ab)V^i_\ab 
 + \frac{1}{15}\oi_\aby V^i_\aby 
 + \frac{1}{105}\hi_\abyd V^i_\abyd\right) \nonumber \\
 &- \frac{1}{2}\sum_i^{n_\mr{MM}}\left(\Delta\mi_\al V^i_\al 
  + \frac{1}{3}\Delta\qi_\ab V^i_\ab\right)
\end{align}
it can be rewritten by using \eqref{eq:selfenergy} to give
\begin{align}
 E^\mr{MM}_\mr{ele+ind} =& \frac{1}{2}\sum^{n_\mr{MM}}_i
 \left((\mi_\al+\Delta\mi_\al) V^i_\al
 + \frac{1}{3}(\qi_\ab + \Delta\qi_\ab) V^i_\ab 
 + \frac{1}{15}\oi_\aby V^i_\aby 
 + \frac{1}{105}\hi_\abyd V^i_\abyd\right) \nonumber \\
 &+ \sum_i^{n_\mr{MM}}\left(\frac{1}{2}\frac{\Delta\mi_\al(\al)\Delta\mi_\be(\al)}{\al_\ab} 
     + \frac{1}{3}\frac{\Delta\mi_\al(\al)\Delta\qi_\by(C)}{k_{\al,\by}}
     + \frac{1}{6}\frac{\Delta\qi_\ab(C)\Delta\qi_{\gamma\delta}(C)}{C_{\gamma\delta,\ab}}\right)\label{eq:trueMM}
\end{align}
At convergence of the MM iterative cycle (equations (2)-(5) in the main text) the following conditions apply
\begin{equation}
    \frac{\partial E^\mr{MM}_\mr{ele+ind}}{\partial \Delta\mi_\al} = 0,\ \frac{\partial E^\mr{MM}_\mr{ele+ind}}{\partial \Delta\qi_\ab} = 0 \label{eq:conditions}
\end{equation}
hence, the self-energy terms vanish when evaluating the explicit forces on the MM COM sites
\begin{align}
 F^i_\al =& -\frac{dE^\mr{MM}_\mr{ele+ind}}{d r^i_\al} \nonumber \\
 =& -\frac{\partial E^\mr{MM}_\mr{ele+ind}}{\partial  r^i_\al} 
    -\frac{\partial E^\mr{MM}_\mr{ele+ind}}{\partial \Delta\mi_\be}
     \frac{\partial \Delta\mi_\be}{\partial  r^i_\al}
    -\frac{\partial E^\mr{MM}_\mr{ele+ind}}{\partial \Delta\qi_\be}
     \frac{\partial \Delta\qi_\be}{\partial  r^i_\al}
\end{align}
and only the first term on the right hand side remains which results in
\begin{equation}
   -\frac{\partial E^\mr{MM}_\mr{ele+ind}}{\partial  r^i_\al} = -\left((\mi_\al+\Delta\mi_\al) V^i_\ab
 + \frac{1}{3}(\qi_\ab + \Delta\qi_\ab) V^i_\aby 
 + \frac{1}{15}\oi_\aby V^i_\abyd 
 + \frac{1}{105}\hi_\abyd V^i_\abyde\right)
\end{equation}
On the other hand the simplified energy expression in equation (12) in the main text -- which gives numerically the same energy as equation (11) in the main text is
\begin{equation}
E^\mr{MM}_\mr{ele+ind} = \frac{1}{2}\sum^{n_\mr{MM}}_i\left(\mi_\al
 V^i_\al
+ \frac{1}{3}\qi_\ab V^i_\ab 
+ \frac{1}{15}\oi_\aby V^i_\aby 
+ \frac{1}{105}\hi_\abyd V^i_\abyd\right) \label{eq:falseMM}
\end{equation}
where a net cancellation is carried out between the self-energy and the intermolecular induced-static and induced-induced electrostatic interactions. This can be misleading. Although these terms are the same in mathematical form and by cancellation result in compact expressions, these terms are not the same in physical origin. Evaluating the forces for this expression results in
\begin{equation}
   -\frac{\partial E^\mr{MM}_\mr{ele+ind}}{\partial  r^i_\al} = -\left(\mi_\al V^i_\ab
 + \frac{1}{3}\qi_\ab V^i_\aby 
 + \frac{1}{15}\oi_\aby V^i_\abyd 
 + \frac{1}{105}\hi_\abyd V^i_\abyde\right)
\end{equation}
which is wrong as the force due to the intermolecular induced-static and induced-induced interaction are not fully accounted for.

The total energy functional expression for the PE-QM/MM interface (equation (53) in the main text) is
\begin{equation}
 E^\mr{sys}_\mr{tot} = E^\mr{KS}[\rho'_\mr{QM}(\br)] + E^\mr{QM/MM}_\mr{ele+ind}[\rho'_\mr{QM}(\br)] +
 E^\mr{MM'}_\mr{ele+ind} + \sum_i^{n_\mr{MM}}\left(E^{\mr{MM},i}_\mr{self} - 
 E^{\mr{MM},i\mr{QM}}_\mr{self}\right) 
 + E^\mr{sys}_\mr{NE} \label{eq:qmmmenergy} .
\end{equation}
It contains two important self-energy terms. One which corresponds to the self-energy due to the total external field and field gradient at site $i$, $E^{\mr{MM},i}_\mr{self}$, and the second which corresponds to the self-energy term excplicitly due to the external field and field gradient from the QM charge density, $E^{\mr{MM},i\mr{QM}}_\mr{self}$. The latter term is subtracted since this term is already present in the QM/MM electrostatic plus induction energy functional term, $E^\mr{QM/MM}_\mr{ele+ind}[\rho'_\mr{QM}(\br)]$.

The expression in equation \eqref{eq:selfenergy} is similarly useful here when evaluating the force on the \ecom\ of the MM sites, since both of the self-energy contributions $E^{\mr{MM},i}_\mr{self}$ and $E^{\mr{MM},i\mr{QM}}_\mr{self}$ can be written solely in terms of the on-site induced moments and polarization matrices. At convergence of QM/MM-SCF, the following conditions apply
\begin{equation}
    \frac{\partial E^\mr{sys}_\mr{tot}}{\partial \Delta\mi_\al} = 0,\ \frac{\partial E^\mr{sys}_\mr{tot}}{\partial \Delta\qi_\ab} = 0,\ \frac{\partial E^\mr{sys}_\mr{tot}}{\partial\ket{\psi_n}} = 0 \label{eq:conditions_qmmm}
\end{equation}
hence these self-energy terms vanish when evaluating the explicit forces on the MM \ecom\ sites 
\begin{align}
 F^i_\al =& -\frac{dE^\mr{sys}_\mr{tot}}{dr^i_\al} \nonumber \\
 =& -\frac{\partial E^\mr{sys}_\mr{tot}}{\partial r^i_\al} 
    -\frac{\partial E^\mr{sys}_\mr{tot}}{\partial \Delta\mi_\be}
     \frac{\partial \Delta\mi_\be}{\partial r^i_\al}
    -\frac{\partial E^\mr{sys}_\mr{tot}}{\partial \Delta\qi_\be}
     \frac{\partial \Delta\qi_\be}{\partial r^i_\al}
    -\frac{\partial E^\mr{sys}_\mr{tot}}{\partial \bra{\psi_m}}
     \frac{\partial \bra{\psi_m}}{\partial r^i_\al} \label{eq:fqmall}
\end{align}
and one only has to consider the first term on the right hand side, which is
\begin{equation}
    -\frac{\partial E^\mr{sys}_\mr{tot}}{\partial r^i_\al} = 
    -\frac{\partial E^\mr{KS}[\rho'_\mr{QM}(\br)]}{\partial r^i_\al}
    -\frac{\partial E^\mr{QM/MM}_\mr{ele+ind}[\rho'_\mr{QM}(\br)]}{\partial r^i_\al}
    -\frac{\partial E^\mr{MM'}_\mr{ele+ind}}{\partial r^i_\al} 
    -\frac{\partial E^\mr{sys}_\mr{NE}}{\partial r^i_\al}
\end{equation}

Applying the same arguments and as shown for the MM energy expressions in equations \eqref{eq:trueMM} and \eqref{eq:falseMM} above, the simplified PE-QM/MM energy expression in equation (55) in the main text leads to wrong forces, while still giving the same numerical energy.



\section{PE-QM/MM Interface Scaling}
\begin{figure}[h!]
\centering
\includegraphics[width=\textwidth]{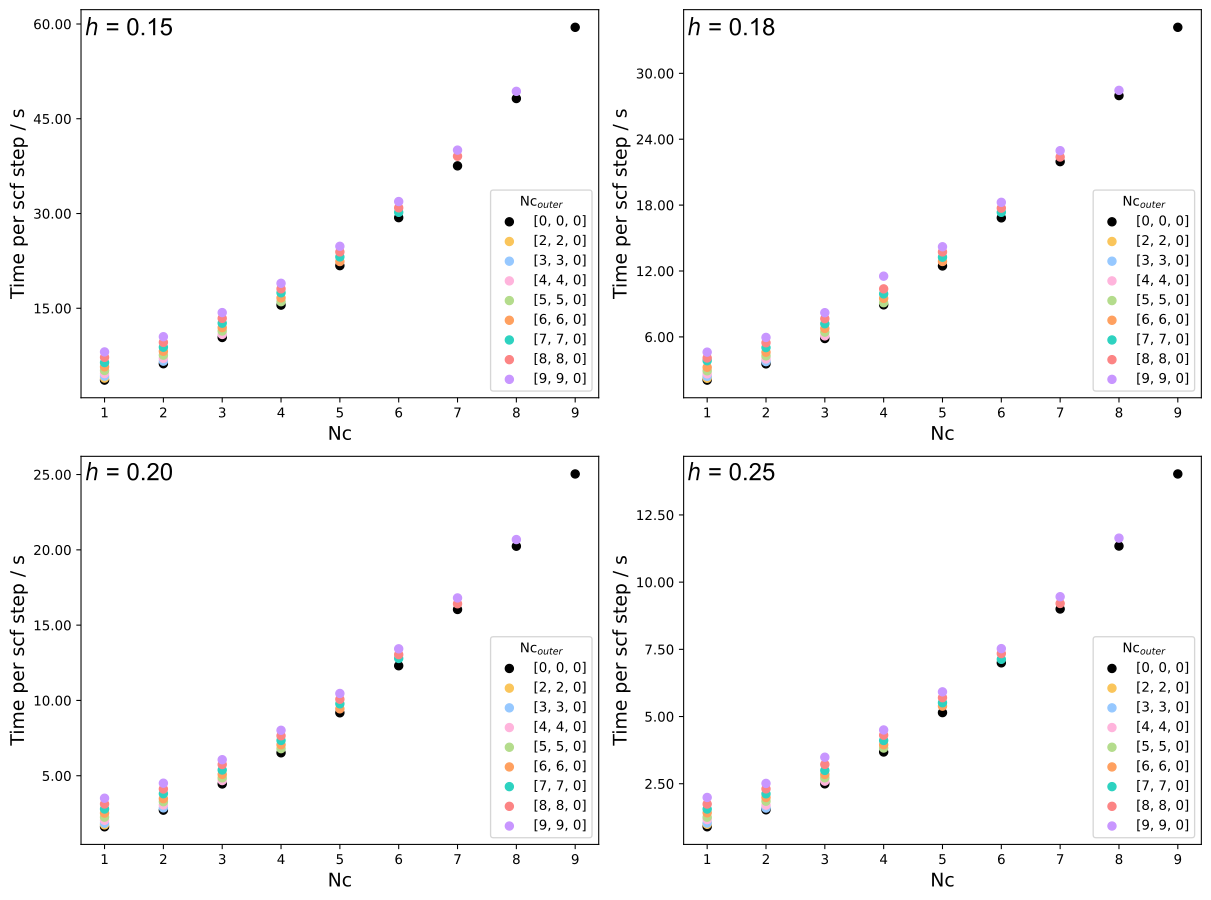}
\caption{Time per scf step calculated for the 2D periodic system of two ice layers for different real-space grids.}
\label{fig:time_vs_Nc}
\end{figure}

\begin{figure}[!ht]
\centering
\includegraphics[width=\textwidth]{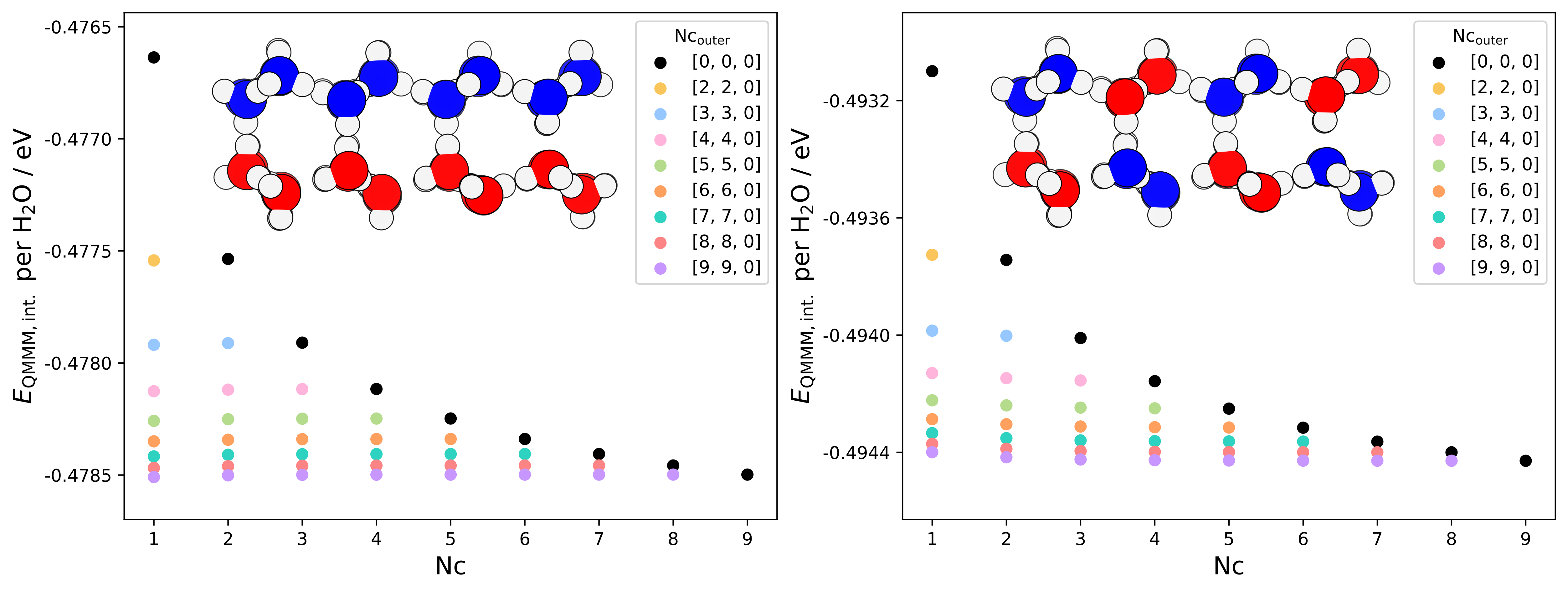}
\caption{QM/MM interaction energy of two ice layers containing a QM (red) and a MM (blue) layer in a 2D periodic system calculated in different cell grid expansion configurations at a real-space grid of $h$ = 0.15.}
\label{fig:E_int_vs_Nc_15}
\end{figure}

\begin{figure}[!ht]
\centering
\includegraphics[width=\textwidth]{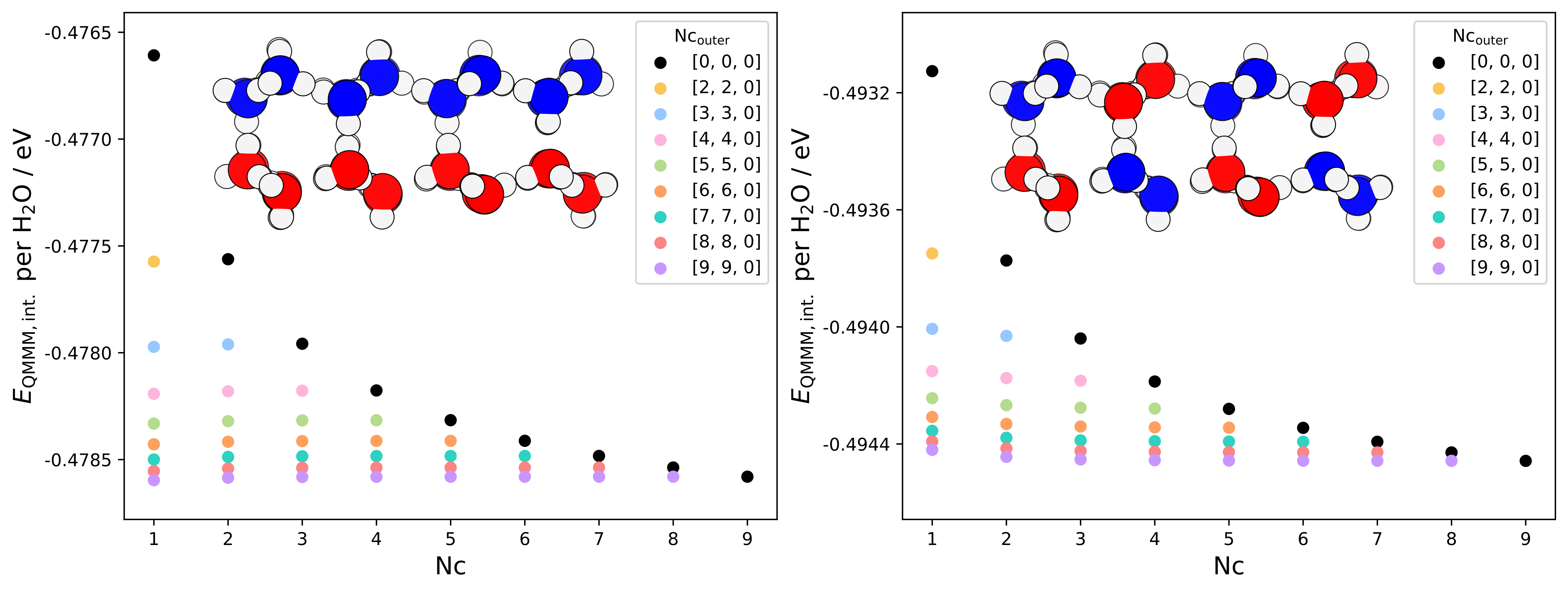}
\caption{QM/MM interaction energy of two ice layers containing a QM (red) and a MM (blue) layer in a 2D periodic system calculated in different cell grid expansion configurations at a real-space grid of $h$ = 0.18.}
\label{fig:E_int_vs_Nc_18}
\end{figure}

\begin{figure}[!ht]
\centering
\includegraphics[width=\textwidth]{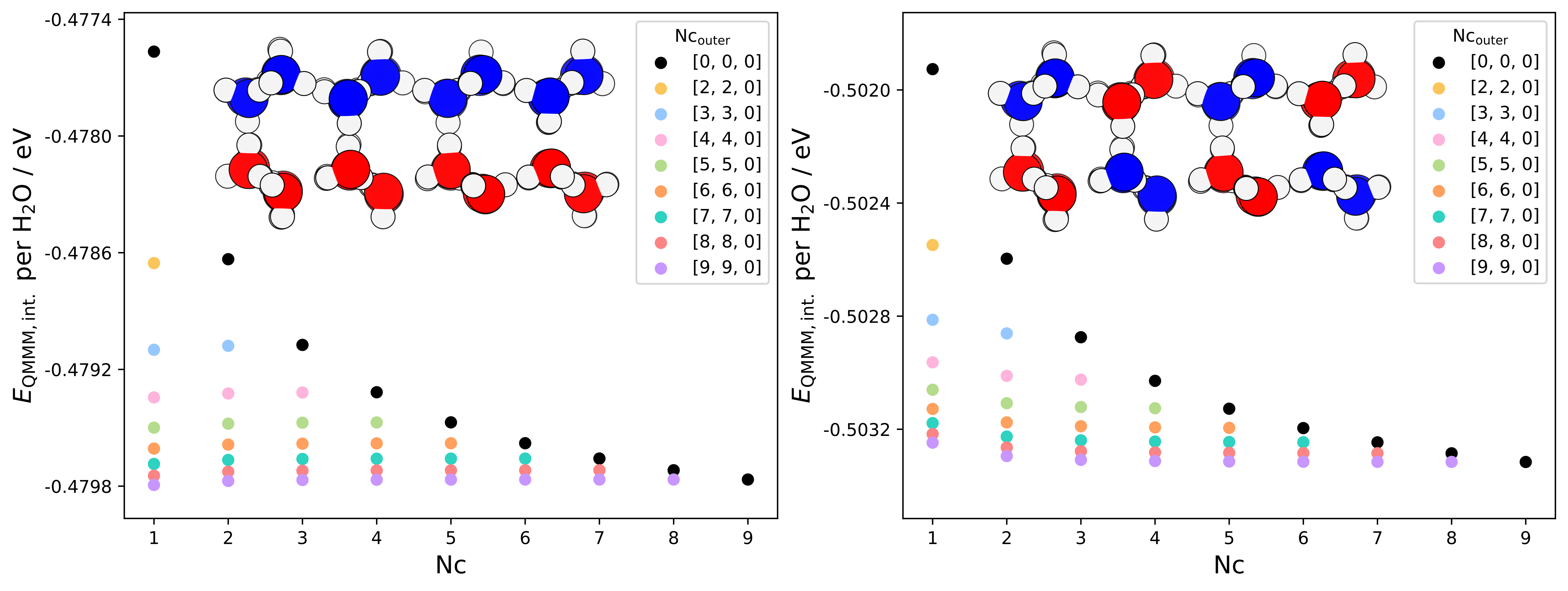}
\caption{QM/MM interaction energy of two ice layers containing a QM (red) and a MM (blue) layer in a 2D periodic system calculated in different cell grid expansion configurations at a real-space grid of $h$ = 0.25.}
\label{fig:E_int_vs_Nc_25}
\end{figure}

The computation time is an important parameter in DFT calculations. In most cases, there is a weigh up between accuracy and speed. To this end, we analyzed the time per SCF step for a PE-QM/MM calculation, as shown in FIG.~\ref{fig:time_vs_Nc}. We tested the real-space grids $h$=0.15, 0.18, 0.20, and 0.25, which showed a similar behavior for the SCF time depending on the $Nc$/$Nc_{\mathrm{outer}}$ cell grid expansion. Increasing the number of inner cell grid $Nc$ leads to a nonlinear increase in time per SCF step. However, the increase in outer cell grids $Nc_{\mathrm{outer}}$ for a given number of $Nc$ adds at most 2.5~s per SCF step, which is less than the increase in $Nc$ expansion. It should be noted that the most accurate outer cell grid expansion of $Nc_{\mathrm{outer}}$ = [9,9,0] is still faster than the increase of $Nc$ by one, starting from $Nc \geq$ [2,2,0]. The additional SCF time for the $Nc_{\mathrm{outer}}$ expansion is negligible compared to the SCF time of a pure $Nc$ calculation.
The QM/MM interaction energy was calculated by 
\begin{equation}
    E_{\mathrm{QMMM, int.}} = \frac{1}{n_{H_2O}} (E_{\mathrm{QMMM, total}} - \sum^{m^{\mathrm{QM}}_{H_2O}}_i E_{\mathrm{QM, i}}),  \label{eq:qmmm_int_energy} 
\end{equation}
where $E_{\mathrm{QMMM, total}}$ defines the total energy of the QM/MM calculation, and $E_{\mathrm{QM, i}}$ is the QM energy of water molecule $i$ which is summed over all water molecules in the QM subsystem of the QM/MM calculation $m^{\mathrm{QM}}_{H_2O}$. Lastly, the energy is normalized by the total number of water molecules in the system $n_{H_2O}$.
Moreover, the QM/MM interaction energy was calculated for the real-space grid $h$ = 0.15, 0.18, 0.20, and 0.25. Comparing FIG.~\ref{fig:E_int_vs_Nc_15}, \ref{fig:E_int_vs_Nc_18}, and \ref{fig:E_int_vs_Nc_25}, the convergence pattern is the same for the layered and mixed systems. The $Nc_{\mathrm{outer}}$ expansion significantly enhances the accuracy for all investigated real-space grids. At $Nc$ = [1,1,0] and $Nc_{\mathrm{outer}}$ = [9,9,0], the interaction energy values are already in agreement with the pure inner cell grid expansion of $Nc$ = [9,9,0].
Together with the analysis of the time per SCF step in FIG.~\ref{fig:time_vs_Nc}, it becomes evident that PE-QM/MM calculations can be accelerated by a factor of five while maintaining the accuracy of an expansive inner cell grid calculation with $Nc$ = [9,9,0]. This underlines the effectiveness of PE-QM/MM simulations, using the multipole expansion to describe long-range electrostatics. 


Next, we compared the QM/MM interaction energy values with the corresponding interaction energy values of the pure QM and pure MM calculations, as reported in table~\ref{tbl:int_E_layered_system} and table~\ref{tbl:int_E_mixed_system}. For that, we calculated the pure QM energy of the total system $E_{\mathrm{QM, total}}$ and the corresponding monomer energy $E_{\mathrm{QM, i}}$ of the water molecule $i$ and extracted the interaction energy using:
\begin{equation}
    E_{\mathrm{QM, int.}} = \frac{1}{n_{H_2O}} (E_{\mathrm{QM, total}} - \sum^{n_{H_2O}}_i E_{\mathrm{QM, i}}).   
\end{equation}
Similarly, we calculated the pure MM energy, using the total MM energy of the ice layer system and dividing it by the number of water molecules.
For a consistent and reliable comparison, we used the most accurate QM/MM interaction energy value, which is the energy value of the $Nc$ = [9,9,0] and $Nc_{\mathrm{outer}}$ = [0,0,0] calculation. In the case of the layered system in table~\ref{tbl:int_E_layered_system}, the calculated interaction energy values of the QM/MM calculation are between the pure QM and pure MM calculations for all real-space grids. However, the values of the QM/MM interaction energy are very close to the values of the pure QM calculation, giving rise to the fact that the PE-QM/MM calculation mimics the pure QM calculation. The exact alignment can be further adjusted by changing the short-range damping value and the isotropic damping value. Moreover, the agreement of the QM/MM interaction energy values with the pure QM energy values is desired, since the SCME has been trained to fit the QM PBE functional. Notably, the QM/MM simulation provides energetics that closely align with the more accurate QM calculation. 
In the case of the mixed system, table~\ref{tbl:int_E_mixed_system} shows that the QM/MM interaction energy values are slightly below the pure QM interaction energy values. The discrepancy is smaller for lower real-space grids. That is, the energetics of PE-QM/MM calculations can exceed the QM energetics for non-layered QM/MM partitionings of the system. 
\begin{table}[!ht]
\centering
\begin{tabular}{lllll}
  \hline
  Real-space grid $h$ & 0.15 & 0.18 & 0.20 & 0.25 \\ 
  \hline
  E$_{\mathrm{MM, int.}}$ / eV & -0.4084 & -0.4084 & -0.4084 & -0.4084\\
  E$_{\mathrm{QMMM, int.}}$ / eV & -0.4785 &  -0.4786 & -0.4787 & -0.4798\\
  E$_{\mathrm{QM, int.}}$ / eV & -0.4837 & -0.4840 & -0.4845 & -0.4873\\
\end{tabular}
\caption{Comparison between the pure QM, pure MM, and QMMM (at $Nc$ = [9,9,0]) interaction energy values per real-space grid of the layered system.}\label{tbl:int_E_layered_system}
\end{table}

\begin{table}[!ht]
\centering
\begin{tabular}{lllll}
  \hline
  Real-space grid $h$ & 0.15 & 0.18 & 0.20 & 0.25 \\ 
  \hline
  E$_{\mathrm{MM, int.}}$ / eV & -0.4084 & -0.4084 & -0.4084 & -0.4084\\ 
  E$_{\mathrm{QMMM, int.}}$ / eV & -0.4944 & -0.4945 & -0.4946 & -0.5169\\
  E$_{\mathrm{QM, int.}}$ / eV & -0.4837 & -0.4840 & -0.4845 & -0.4873\\
\end{tabular}
\caption{Comparison between the pure QM, pure MM, and QMMM (at $Nc$ = [9,9,0]) interaction energy values per real-space grid of the mixed system.}\label{tbl:int_E_mixed_system}
\end{table}
\clearpage

\section{Finding clusters and shifting multipole moments}
\subsection{K-means clustering}
The K-means algorithm partitions a set of $N$ data points
$\{\vec{x}_1, \ldots, \vec{x}_N\}$ into $K$ disjoint clusters by
iteratively minimising the total within-cluster variance.
Each data point $\vec{x}_n$ is a vector of coordinates (here, the
three-dimensional position of an MM site).
The algorithm maintains $K$ \emph{centroids}
$\{\vec{\mu}_1, \ldots, \vec{\mu}_K\}$, where centroid $\vec{\mu}_k$
is the geometric mean of all points currently assigned to cluster $k$.
Cluster membership for each point is encoded by the set $\omega_k$,
which collects every point closest to centroid $\vec{\mu}_k$.
Initialisation follows the k-means$++$ strategy
\cite{Arthur2007kmeansPP}, which spreads the initial centroids across the data
to accelerate convergence and avoid poor local minima.
The algorithm alternates between two steps: reassignment and
centroid update until, the centroid positions change by less than a
chosen tolerance $\varepsilon$ (or a maximum number of iterations is
reached), the version of this algorithm as implemented was used in this publication
\texttt{sklearn.cluster.KMeans}.

\begin{figure}[!ht]
\centering
\includegraphics[width=\textwidth]{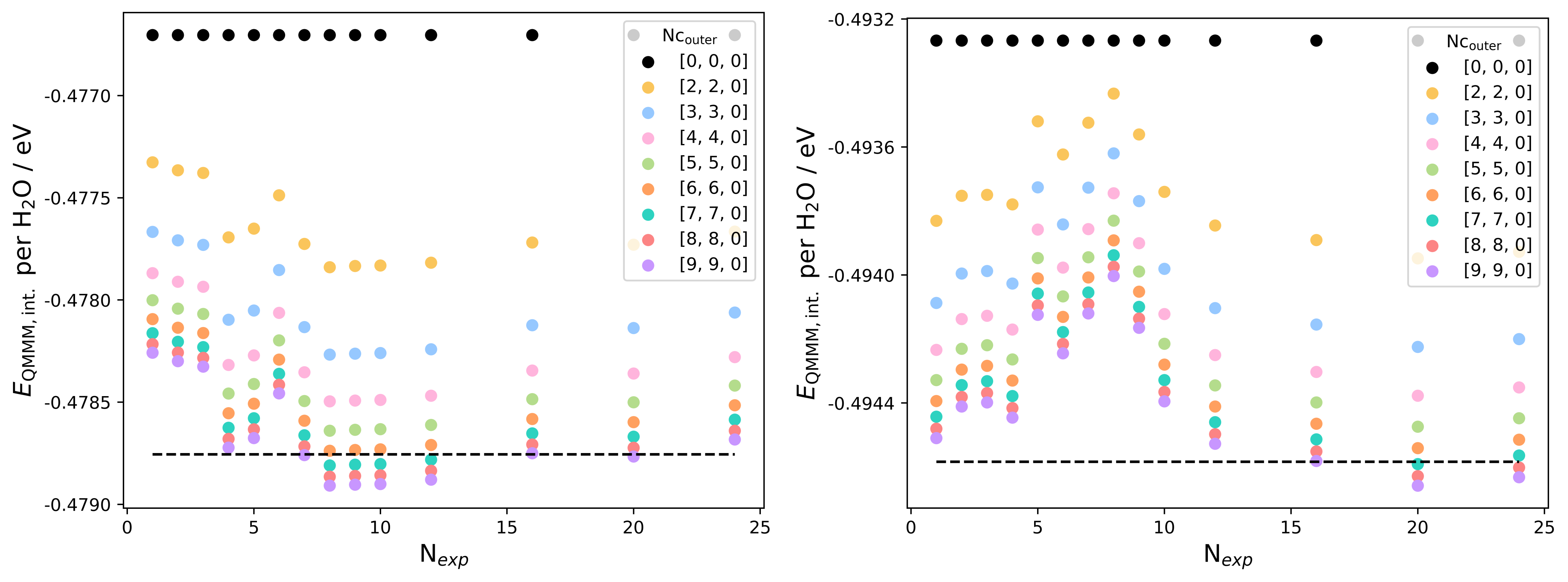}
\caption{QM/MM interaction energy of two ice layers with a layered QM/MM partitioning (left) and a mixed QM/MM partitioning (right) in a 2D periodic system calculated in $Nc$ = [1,1,0] and $Nc_{\mathrm{outer}}$ = [9,9,0], a real-space grid of $h$ = 0.20, and at expansion points $N_{\mathrm{exp}}$ from 1 to 24, generated by the K-means algorithm. Moreover, the dashed line illustrates the converged interaction energy with an inner cell grid expansion of $Nc$ = [9,9,0].}
\label{fig:E_int_vs_Nexp}
\end{figure}

Using the K-means algorithm, we investigated the influence of the number of centroids on the accuracy regarding the QM/MM interaction energy values. The distribution of the energy values is different for the layered and mixed QM/MM ice layer system. The layered system, illustrated in the left panel in FIG.~\ref{fig:E_int_vs_Nexp}, suggests that dividing the system into more clusters results in a better accuracy for a low $Nc$ expansion. The distribution of the energy values for different $Nc_{\mathrm{outer}}$ stays the same for different numbers of expansion centroids. Thus, the change of the number of expansion centroids leads to a uniform shift for all $Nc_{\mathrm{outer}}$ configurations. 
In the case of the mixed ice layer system, the increase in the number of centroids does benefit the accuracy. The interaction energy values start to increase up to the maximum difference of 0.0005 eV per \ce{H2O} from the convergence value at eight centroids. Then, the energy values decrease again and converge to the convergence value of the $Nc$ = [9,9,0] calculation. 
Thus, the partitioning of the QM/MM system matters, as reducing the number of centroids can worsen the accuracy of the QM/MM calculation. 
Currently, the K-means algorithm is implemented only on the basis of the geometric vicinity of the solvent molecule's center of mass. Thus, the clustering can be significantly different when the QM/MM configuration is changed. Thereby, four centroids are in good agreement with the convergence value in the layered system, whereas four centroids are not beneficial for the mixed ice layer system.
Beyond that, the increase and decline at eight centroids in the mixed system can be explained by the dipole magnitude in the clusters. Although the water molecules are spatially close to each other in the cluster, the dipole contribution of that cluster might be completely different from that of other clusters. We think that spatial clustering could lead to an increased dipole moment value for some clusters, exceeding the total dipole moment of the cell, which impacts their electrostatic contribution to the origin-shifting on the outer cell grid. Therefore, the partitioning exhibits an unexpected convergence pattern for the interaction energy in some QM/MM configurations. To this end, the impact of shifting the multipole moments has to be further investigated.

\begin{algorithm}[H]
\DontPrintSemicolon
\SetAlgoLined
\SetKwComment{Comment}{\textit{(}}{)}

\caption{K-\textsc{Means}$\bigl(\{\vec{x}_1,\ldots,\vec{x}_N\},\,K\bigr)$}

\KwIn{%
  $\{\vec{x}_1, \ldots, \vec{x}_N\}$: set of $N$ data points;
  $K$: number of clusters}
\KwOut{%
  $\{\vec{\mu}_1, \ldots, \vec{\mu}_K\}$: final cluster centroids}

\BlankLine
\tcp*[l]{\textbf{Initialisation} --- place centroids using k-means++}
Initialise $K$ centroids $\vec{\mu}_1, \ldots, \vec{\mu}_K$
from $\{\vec{x}_1, \ldots, \vec{x}_N\}$ via k-means$++$\;

\BlankLine
\tcp*[l]{\textbf{Iterate} until convergence}
\While{centroids change by more than tolerance $\varepsilon$}{

    \BlankLine
    \tcp*[l]{--- Reassignment step ---}
    \For{$k \leftarrow 1$ \KwTo $K$}{
        $\omega_k \leftarrow \{\}$
        \Comment*[r]{empty each cluster}
    }
    \For{$n \leftarrow 1$ \KwTo $N$}{
        $k^* \leftarrow \arg\min_{k}\;\|\vec{x}_n - \vec{\mu}_k\|$
        \Comment*[r]{find nearest centroid}
        $\omega_{k^*} \leftarrow \omega_{k^*} \cup \{\vec{x}_n\}$
        \Comment*[r]{assign point to that cluster}
    }

    \BlankLine
    \tcp*[l]{--- Update step ---}
    \For{$k \leftarrow 1$ \KwTo $K$}{
        $\vec{\mu}_k \leftarrow \dfrac{1}{|\omega_k|}
          \displaystyle\sum_{\vec{x}\,\in\,\omega_k} \vec{x}$
        \Comment*[r]{new centroid = mean of cluster}
    }
}

\BlankLine
\KwRet{$\{\vec{\mu}_1, \ldots, \vec{\mu}_K\}$}

\end{algorithm}

\vspace{1cm}
After clustering the \ce{H2O} molecules, the cluster centers are used as a coarse grid onto which the multipole moments are shifted, yielding a more accurate representation of the electrostatics while remaining computationally efficient. The equations used for shifting the moments in cartesian coordinates is given in the following sections.

\subsection{Multipole Origin Shift}
General expressions for shifted dipole up to and including hexadecapole are presented below. In general this transform multipole moment $M_i\rightarrow M'_i$, of expansion center $i$. In all cases the multipoles are shifted by the common vector 
\begin{equation}
 \mathbf{R} = \mathbf{R}_\mathrm{new} - \mathbf{R}_{i}
\end{equation}
where $\mathbf{R}_\mathrm{new}$ is the new origin -- and is possibly common to multiple origin shifted expansion centers -- and $\mathbf{R}_{i}$ is the origin of expansion center $i$. 

The total shifted multipole moment at the new origin is then simply given by
\begin{equation}
 M'_\mathrm{tot} = \sum_iM'_i
\end{equation}

\subsection{Dipole-shift}

Dipole vector component $\mu^o_{i}$ of a system with charge $q$, when shifted by the vector $\mathbf{R} = \{R_x,R_y,R_z\}$ results in $\mu'_{i}$ given by: 
\begin{align}
    \mu^{'}_{x} &= \mu^{o}_{x} - qR_x \\
    \mu^{'}_{y} &= \mu^{o}_{y} - qR_y \\
    \mu^{'}_{z} &= \mu^{o}_{z} - qR_z 
\end{align}

\textbf{Note:} In a neutral system ($q=0$), the dipole moment is origin-independent.

\subsection{Quadrupole-shift} 
Quadrupole tensor component $\Theta^o_{i,j \in \{x,y,z\}}$ of a system with charge $q$, and dipole moment $\mu^o_{ k\in \{x,y,z\} }$ when shifted by the vector $\mathbf{R} = \{R_x,R_y,R_z\}$, results in $\Theta'_{i,j}$ given by: 

\begin{align}
    \Theta^{'}_{x,x} &= \Theta^{o}_{x,x} -2\mu^{o}_{x}R_x + \mu^{o}_{y}R_y + \mu^{o}_{z}R_z +  q(R_x^2 -\frac{1}{2}(R_y^2 + R_z^2)) \\
    \Theta^{'}_{x,y} &= \Theta^{o}_{x,y} -\frac{3}{2}(\mu^{o}_{x}R_y + \mu^{o}_{y}R_x) + \frac{3}{2}qR_xR_y \\
    \Theta^{'}_{x,z} &= \Theta^{o}_{x,z} -\frac{3}{2}(\mu^{o}_{x}R_z + \mu^{o}_{z}R_x) + \frac{3}{2}qR_xR_z \\
    \Theta^{'}_{y,y} &= \Theta^{o}_{y,y} -2\mu^{o}_{y}R_y + \mu^{o}_{x}R_x + \mu^{o}_{z}R_z + q(R_y^2 -\frac{1}{2}(R_x^2 + R_z^2)) \\
    \Theta^{'}_{y,z} &= \Theta^{o}_{y,z} -\frac{3}{2}(\mu^{o}_{y}R_z + \mu^{o}_{z}R_y) + \frac{3}{2}qR_yR_z \\
    \Theta^{'}_{z,z} &= \Theta^{o}_{z,z} -2\mu^{o}_{z}R_z + \mu^{o}_{y}R_y + \mu^{o}_{x}R_x +  q(R_z^2 -\frac{1}{2}(R_y^2 + R_x^2)) \\
\end{align}

\subsection{Octapole-shift} 
Octapole tensor component $\Omega^o_{i,j,l \in \{x,y,z\}}$ of a system with charge $q$, dipole moment $\mu^o_{ k\in \{x,y,z\} }$ and $\Theta^o_{m,n \in \{x,y,z\}}$ when shifted by the vector $\mathbf{R} = \{R_x,R_y,R_z\}$ results in $\Omega'_{i,j,l}$ given by: 
\begin{align}
    \Omega^{'}_{xxx} &= \Omega^{o}_{xxx} -3\Theta^{o}_{xx}R_{x} +2\Theta^{o}_{xy}R_{y} +2\Theta^{o}_{xz}R_{z} \nonumber \\
    &+ \mu^{o}_{x}(3R^{2}_{x}-\frac{3}{2}R^{2}_{y} -\frac{3}{2}R^{2}_{z}) -3\mu_{y}R_{x}R_{y} -3\mu_{z}R_{x}R_{z} \nonumber\\
    &+q(-R^{3}_{x} +\frac{3}{2}R_{x}R^{2}_{y}+\frac{3}{2}R_{x}R^{2}_{z})\\
    \Omega^{'}_{xxy} &= \Omega^{o}_{xxy} +(\frac{2}{3}\Theta^{o}_{yy} -\frac{5}{3}\Theta^{o}_{xx})R_{y} - \frac{8}{3}\Theta^{o}_{xy}R_{x} \nonumber\\
    &+\frac{2}{3}\Theta^{o}_{yz}R_{z} +\mu^{o}_{y}(-\frac{3}{2}R^{2}_{y} + 2R^{2}_{x} - \frac{1}{2}R^{2}_{z}) + 4\mu_{y}R_{x}R_{y} \nonumber \\
    & -\mu_{z}R_{y}R_{z} +q(\frac{1}{2}R^{3}_{y} -2R_{y}R^{2}_{x}+\frac{1}{2}R_{y}R^{2}_{z})\\
    \Omega^{'}_{xyz} &= \Omega^{o}_{xyz} -\frac{5}{3}\Theta^{o}_{yz}R_{x} -\frac{5}{3}\Theta^{o}_{xz}R_{y} -\frac{5}{3}\Theta^{o}_{xy}R_{z}  \nonumber \\
    & + \frac{5}{2}\mu^{o}_{z}R_x R_y + \frac{5}{2}\mu^{o}_{x}R_y R_z + \frac{5}{2}\mu^{o}_{y}R_z R_x  -\frac{5}{2}qR_xR_yR_z 
\end{align}

\textbf{Note:} This is the irreducible set of equations, and other components can be obtained by replacing the components (say x with y or z) and taking permutations. 
\subsection{Hexadecapole-shift} 
Hexadecapole tensor component $\phi^o_{i,j,k,l \in {x,y,z}}$ of a system with charge $q$, dipole moment $\mu^o_{ m\in \{x,y,z\} }$, quadrupole moment $\Theta^o_{n,o \in \{x,y,z\}}$  and octapole moment  $\Omega^o_{p,q,r \in {x,y,z}}$ when shifted by the vector $\mathbf{R} = \{R_x,R_y,R_z\}$ results in  $\phi'_{i,j,k,l}$ given by:

\begin{align}
\Phi'_{xxxx}
    &= \Phi^0_{xxxx}
     - 4\Omega_{xxx}R_x
     + 3\Omega_{yxx}R_y
     + 3\Omega_{zxx}R_z
     + 6\theta_{xx}R_x^2 \nonumber\\
    &+ \left(-\frac{5}{2}\theta_{xx} + \theta_{yy}\right)R_y^2
     + \left(-\frac{5}{2}\theta_{xx} + \theta_{zz}\right)R_z^2
     - 8\theta_{xy}R_x R_y
     + 2\theta_{yz}R_y R_z \nonumber\\
    &- 8\theta_{xz}R_x R_z
     - 4\mu_x R_x^3
     - \frac{3}{2}\mu_y R_y^3
     - \frac{3}{2}\mu_z R_z^3
     + 6\mu_y R_x^2 R_y \nonumber\\
    &- \frac{3}{2}\mu_z R_y^2 R_z
     + 6\mu_x R_x R_z^2
     + 6\mu_x R_x R_y^2
     - \frac{3}{2}\mu_y R_y R_z^2
     + 6\mu_z R_x^2 R_z \nonumber\\
    &+ q\!\left(
        R_x^4
        + \frac{3}{8}R_y^4
        + \frac{3}{8}R_z^4
        - 3R_x^2 R_y^2
        + \frac{3}{4}R_y^2 R_z^2
        - 3R_x^2 R_z^2
      \right)
\end{align}

\begin{align}
\Phi'_{xxyy} &= \Phi^0_{xxyy}
    + \left(\frac{1}{2}\Omega_{xxx} - 3\Omega_{xyy}\right)R_x
    + \left(\frac{1}{2}\Omega_{yyy} - 3\Omega_{yxx}\right)R_y
    - \frac{1}{2}\Omega_{zzz}R_z \nonumber \\ 
    &+ \left(-\frac{7}{4}\theta_{xx} + \frac{5}{2}\theta_{yy}\right)R_x R_x
    + \left(-\frac{7}{4}\theta_{yy} + \frac{5}{2}\theta_{xx}\right)R_y R_y
    + \frac{3}{4}\theta_{zz}R_z R_z  \nonumber \\ 
    &+ 9\theta_{xy}R_x R_y 
    - \theta_{yz}R_y R_z
    - \theta_{xz} R_x R_z
    + 2\mu_x R_x R_x R_x
    + 2\mu_y R_y R_y R_y \nonumber \nonumber \\ 
    & - \frac{1}{2}\mu_z R_z R_z R_z 
    - \frac{27}{4}\mu_y R_x R_x R_y
    + \frac{3}{4}\mu_z R_y R_y R_z
    + \frac{3}{4}\mu_x R_x R_z R_z \nonumber \\
    & - \frac{27}{4}\mu_x R_x R_y R_y 
    + \frac{3}{4}\mu_y R_y R_z R_z
    + \frac{3}{4}\mu_z R_z R_x R_x  \nonumber \\
    &+ q \left(
        -\frac{1}{2} R_x^4
        - \frac{1}{2} R_y^4
        + \frac{1}{8} R_z^4
        + \frac{27}{8} R_x^2 R_y^2
        - \frac{3}{8} R_y^2 R_z^2
        - \frac{3}{8} R_x^2 R_z^2
    \right)
\end{align}

\begin{align}
\Phi'_{xxxy}
    &= \Phi^0_{xxxy}
     - \frac{15}{4}\Omega_{yxx}R_x
     + \left(-\frac{7}{4}\Omega_{xxx} + \frac{3}{2}\Omega_{xyy}\right)R_y
     + \frac{3}{2}\Omega_{xyz}R_z \nonumber\\
    &+ \theta_{xy}\!\left(5R_x^2 - \frac{15}{4}R_y^2 - \frac{5}{4}R_z^2\right)
     + \left(\frac{35}{4}\theta_{xx} + \frac{5}{2}\theta_{zz}\right)R_x R_y
     - \frac{5}{2}\theta_{xz}R_y R_z \nonumber\\
    &- \frac{5}{2}\theta_{yz}R_x R_z
     - \frac{5}{2}\mu_y R_x^3
     + \frac{15}{8}\mu_x R_y^3
     - \frac{15}{2}\mu_x R_x^2 R_y \nonumber\\
    &+ \frac{15}{8}\mu_y R_z^2 R_x
     + \frac{45}{8}\mu_y R_x R_y^2
     + \frac{15}{8}\mu_x R_y R_z^2
     + \frac{15}{4}\mu_z R_x R_y R_z \nonumber \\
     &+ q (\frac{5}{2} R_x^3 R_y
            - \frac{15}{8} R_x R_y^3
            - \frac{15}{8} R_z^2 R_x R_y)
\end{align}

\begin{align}
\Phi'_{zzxy}
    &= \Phi^0_{zzxy}
     + \left(-\frac{7}{4}\Omega_{zzy} + \frac{1}{2}\Omega_{yxx}\right)R_x
     + \left(-\frac{7}{4}\Omega_{zzx} + \frac{1}{2}\Omega_{xyy}\right)R_y
     - 3\Omega_{xyz}R_z \nonumber\\
    &+ \theta_{xy}\!\left(-\frac{5}{4}R_x^2 - \frac{5}{4}R_y^2 + \frac{5}{2}R_z^2\right)
     + \frac{15}{4}\theta_{zz}R_x R_y
     + 5\theta_{zx}R_y R_z
     + 5\theta_{yz}R_z R_x \nonumber\\
    &+ \frac{5}{8}\mu_y R_x^3
     + \frac{5}{8}\mu_x R_y^3
     + \frac{15}{8}\mu_x R_x^2 R_y
     - \frac{15}{4}\mu_y R_z^2 R_x \nonumber\\
    &+ \frac{15}{8}\mu_y R_x R_y^2
     - \frac{15}{4}\mu_x R_y R_z^2
     - \frac{15}{2}\mu_z R_x R_y R_z \nonumber \\
     &+ q(
     -\frac{5}{8} R_x^3 R_y 
     -\frac{5}{8} R_x R_y^3
     + \frac{15}{4} R_z^2 R_x R_y
     )
\end{align}

\textbf{Note:} This is the irreducible set of equations, and other components can be obtained by replacing the components (say x with y or z) and taking permutations. 

\section{Isotropic Damping}
\begin{figure}[!ht]
\centering
\includegraphics[width=\textwidth]{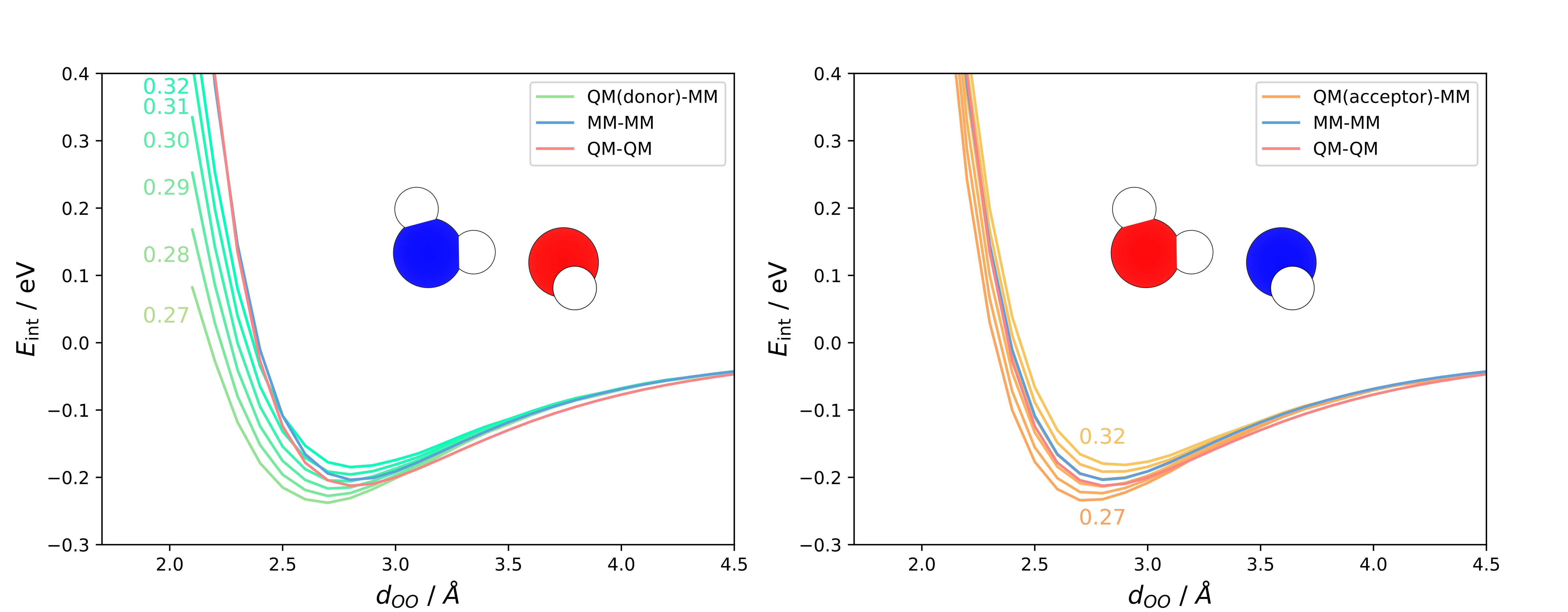}
\caption{Dimer binding energy curve of the pure QM and MM water dimer, and the QM(don.)MM (left) and QM(acc.)MM (right) water dimer, respectively. The QMMM dimer binding curves are shown for various values of $\beta$ from 0.27 up to 0.32.}
\label{fig:dimer_binding_curve}
\end{figure}

\begin{figure}[!ht]
\centering
\includegraphics[width=\textwidth]{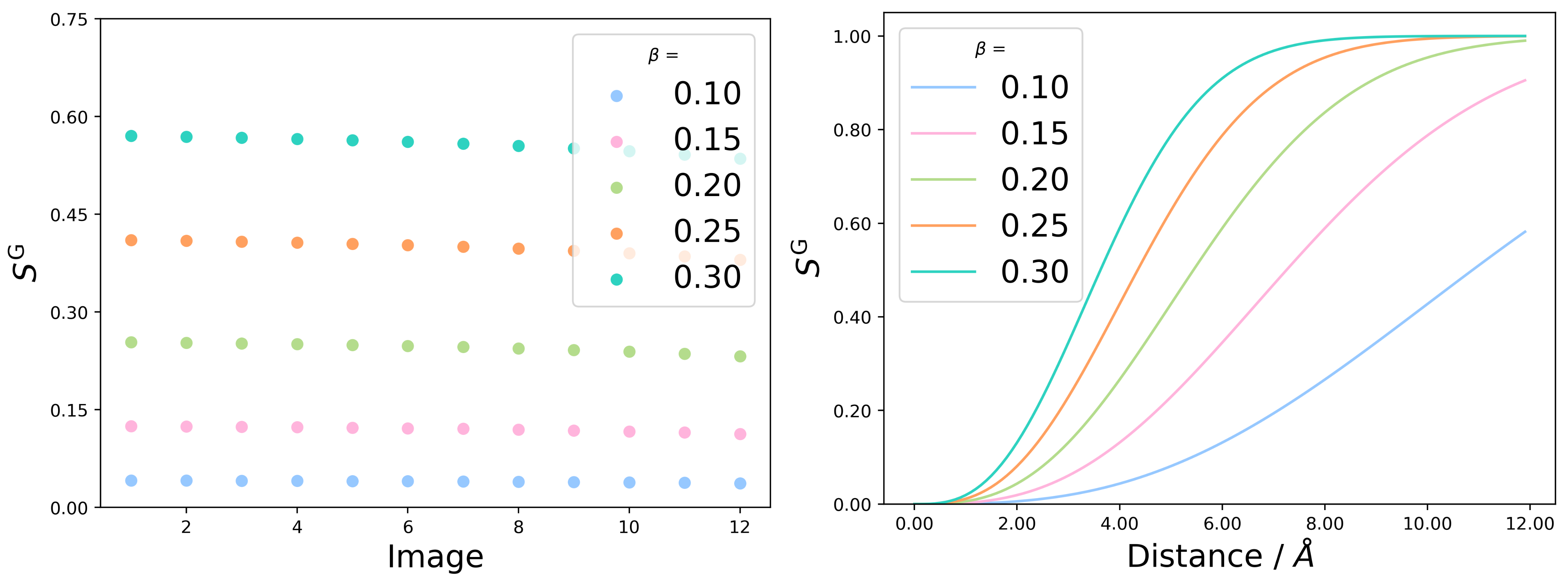}
\caption{$S^G$ values of a dimer interaction 12 images before the polarization catastrophe in the Au water QM/MM MD simulation for different values of $\beta$.}
\label{fig:MD_blow_up_S}
\end{figure}

To handle the polarization catastrophe in the PE-QM/MM simulations, we introduced the isotropic damping (see main text eq. (25)) for solvent molecules at the QM/MM boundary. The isotropic damping and the influence of the parameter $\beta$ are presented in FIG.~\ref{fig:dimer_binding_curve} for a water dimer. 
The values of $\beta$ are positioned close to the corresponding curves. Thus, a smaller value of $\beta$ leads to a lower minimum, which is also shifted to a smaller O-O distance of 2.6\ ~\AA. The influence of the isotropic damping is significant for the QM/MM configuration, with the QM molecule being the donor and the MM molecule being the acceptor. The optimal value of $\beta$ has to be chosen such that the dimer binding energy curve is between the pure QM and pure MM energy curves. Especially, in the area of the minimum, otherwise the QM/MM interaction exhibits inaccuracies at the boundary.  In addition to that, the slope of the dimer binding energy curve in the range of 2.2 to 2.6~\AA is not as steep for lower values of $\beta$. Hence, the repulsive interaction begins at smaller O-O distances, leading to an unphysical distance between water molecules. Eventually, an inadequate choice of $\beta$ can lead to an over-polarization, resulting in the MD blow up, described in the main text of this article. The influence of $\beta$ is not that strong in the case with the QM being the acceptor and the MM being the donor. Here, lowering $\beta$ results in a lower minimum as well, but the repulsive behavior is quite similar for all values of $\beta$ between 0.27 and 0.32.
Eventually, we analyzed the value of $\beta$ during the gold water PE-QM/MM MD simulation before the MD blow up. FIG.~\ref{fig:MD_blow_up_S} illustrates the values of the isotropic damping function of a chosen QM/MM dimer at the boundary with the corresponding $\beta$ for each image. On the right side, the value of $S^G$ is shown as a function of distance with the corresponding $\beta$. 
The impact of the isotropic damping can be magnified as the $\beta$ is smaller. Reducing the value of $\beta$ to 0.10 restricts the electrostatic interaction between a QM and MM molecule to only 5~\%. The impact of such a high damping is presented on the right, as the electrostatic interaction between the QM and MM solvent molecule is reduced to ca. 60~\% even at a distance of 12~\AA. For $\beta$ = 0.30, the isotropic damping stops at a distance of 6~\AA.
Therefore, the choice of $\beta$ depends not only on the diminished electrostatic interaction but also on the range of its effect.
The isotropic damping value $S^G$ stays constant for all values of $\beta$ along the 12 images. This shows that the water dimer is locked in place already at 12 steps before the polarization catastrophe.

\clearpage
\bibliography{main}